\documentclass[twocolumn]{aa} 
\usepackage{graphicx}
\usepackage{txfonts}
\usepackage{natbib}
\bibpunct{(}{)}{;}{a}{}{,} 

\def\gtrsim{\mathrel{\hbox{\rlap{\hbox{\lower4pt\hbox{$\sim$}}}\hbox{$>$}}}}
\def\lesssim{\mathrel{\hbox{\rlap{\hbox{\lower4pt\hbox{$\sim$}}}\hbox{$<$}}}}
\newcommand{\hi}{H{\sc i} }
\newcommand{\kms}{km s$^{-1}$}
\newcommand{\msun}{M$_{\odot}$$\,$}

\newcommand{\cmsq}{cm$^{-2}$$\,$}

\newcommand{\ml}{M$_{HI}$/L$_B$$\,$}
\title{The \hi content of early-type galaxies from the ALFALFA survey}
\subtitle{II. The case of low density environments}
  \author{M. Grossi
          \inst{1}
          \and
          S. di Serego Alighieri\inst{1} \and C. Giovanardi\inst{1} \and
           G. Gavazzi\inst{2} \and R. Giovanelli\inst{3} \and M.P. Haynes\inst{3} \and \\ B.R.Kent\inst{4} \and S. Pellegrini\inst{5} \and S. Stierwalt\inst{3} \and G. Trinchieri\inst{6} }

\institute{INAF-Osservatorio Astrofisico di Arcetri, L.go E. Fermi 5, 50125 Florence, IT\\
\email{grossi@arcetri.astro.it}
\and Universit\'a di Milano-Bicocca, Piazza delle Scienze 3, 20126 Milan, IT
\and Center for Radiophysics and Space Research, Cornell University, Ithaca, NY 14853
\and Jansky Fellow of the National Radio Astronomy Observatory, 520 Edgemont Road,
Charlottesville, VA 22903.
\and Universit\'a di Bologna, via Ranzani 1, 40127, Bologna, IT
\and INAF-Osservatorio Astronomico di Brera, via Brera 28, 20121, Milan, IT}

\begin{document}

\date{}


{}

\abstract{}{We present the analysis of the \hi content of a sample of early-type galaxies (ETGs) in low-density environments (LDEs) using the data set provided by the Arecibo Legacy Fast ALFA (ALFALFA) survey. We compare their properties to the sample in the Virgo cluster that we studied in a previous paper (di Serego Alighieri et al. 2007, Paper I). Our aim is both to investigate how the cool interstellar medium (ISM) of these systems depends on the galaxy mass and the environment and to relate the properties of the neutral hydrogen to the warm
phases of the ISM.}{We have selected a sample of 62 nearby ETGs (V$<$ 3000 \kms) in an area of the sky where the ALFALFA data are already available ($8^h<$RA$<16^h$, $4^{\circ}<$DEC$<16^{\circ}$), avoiding the region of the Virgo cluster. Among these, 39 have absolute B magnitudes fainter than M$_B = -17$.
}{Fifteen out of 62 galaxies have been firmly detected with ALFALFA  ($\sim$ 25\%). Five additional galaxies show a weaker \hi emission (S/N $\sim 4$) and they will need deeper observations to be confirmed.
Eight objects had 21-cm measurements reported in the literature. One by one comparison with the available material
confirms, as expected, that ALFALFA data are,
with rare exceptions, of equal or better quality than the best spectra
previously obtained for these objects.
All together, our analysis doubles the number of known gas-rich ETGs in this area.
The \hi detection rate is 44\% in luminous ETGs (M$_B < -17$) and 13\% in dwarf ETGs (M$_B > -17$). In both cases it is 10 times higher than that of the Virgo cluster.
The presence of gas
can be related to a recent star formation activity:  60\% of all ETGs with \hi have  optical emission line ratios typical of star-forming galaxies and blue colours suggesting the presence of young stellar populations, especially in the dwarf subsample.}{We show that the \hi detection rate of ETGs depends both on the environment and mass. The fraction of early-type systems with neutral hydrogen is higher in more massive objects when compared to early-type dwarfs. The ETGs in LDEs seem to have more heterogeneous properties than their Virgo cluster counterparts, since they are able to  retain  a cold interstellar gas component and to support star formation activity even at recent epochs.
}


\maketitle

\section{Introduction}

\begin{table*}
\caption{The sample of early-type galaxies selected in this work and their main parameters.
}
\begin{center}
\begin{minipage}{\textwidth}
\begin{tabular}{lcrcrcccrrrc}
\hline \hline
 ID            &  M$_B$ &V$_{opt}$  &V$_{HI}$$\;\;\,$ &$\Delta$V$_{50}$    &  $F$    & M$_{HI}$    & \small log$\frac{M_{HI}}{L_B}$ & m-M & \tiny H${\alpha}$ &   T &\small code \\
         & \small  mag      & \small \kms  & \small \kms   & \small  \kms   & \small Jy \kms       &  \small 10$^7$\msun& \small  & \small  mag$\,$ & \small  \AA$\,$      &  & \\
\hline \hline
                UGC4590 & -17.26 & 1901 &    -- & 208 &  $<$0.65 &  $<$13.3 & $<$-0.97 & 32.17 &  11.3 &  -1 & -- \\
                UGC4599 & -19.07 & 2072 & 2071$\pm$ 1 & 148 & 35.34$\pm$0.09 & 850.5 &  0.11 & 32.53 &  -0.01 &  -2 & $a$ \\
SDSSJ093155.94+073210.3 & -14.30 & 2217 &    -- &  83 &  $<$0.42 &  $<$11.7 &  $<$0.16 & 32.49 &  -1.7 &  -2 & -- \\
SDSSJ093208.81+082810.8 & -14.41 & 2340 &    -- &  83 &  $<$0.46 &  $<$14.3 &  $<$0.20 & 32.60 &  -3.6 &  -5 & -- \\
                MRK0706 & -17.56 & 2489 & 2501$\pm$ 7 &  61 &  0.64$\pm$0.06 &  22.5 & -0.86 & 32.73 & 123.0 &  -1 & $b$ \\
SDSSJ093608.59+061525.4 & -15.66 & 2424 & 2413$\pm$11 & 127 &  0.60$\pm$0.06 &  19.7 & -0.16 & 32.66 &  21.1 &  -1 & $b$ \\
                NGC2962 & -20.08 & 1959 & 1958$\pm$ 1 & 414 &  4.22$\pm$0.12 & 112.7 & -1.17 & $^{\star}$32.64 & -- &  -1 & $a$ \\
SDSSJ094218.94+044121.8 & -13.67 & 2036 &    -- &  83 &  $<$0.50 &  $<$12.1 &  $<$0.42 & 32.31 &   0.5 &  -5 & -- \\
SDSSJ095036.25+124832.7 & -14.49 & 1335 &    -- &  83 &  $<$0.41 &   $<$4.8 & $<$-0.31 & 31.54 &   9.3 &  -5 & -- \\
            CGCG064-021 & -18.48 & 2822 & 2676$\pm$ 9 &  36 &  0.30$\pm$0.07 &  13.1 & -1.47 & 32.96 &  -1.0 &  -5 & $c$ \\
            CGCG064-055 & -17.92 & 2789 &    -- & 210 &  $<$0.68 &  $<$30.2 & $<$-0.88 & 32.99 &  20.7 &  -5 & -- \\
                   LeoI & -11.00 &  229 &    -- &  82 &  $<$0.51 &   $<$6.3$\times10^{-4}$ & $<$-2.79 & $^{\star}$21.80 & -- &  -5 & -- \\
SDSSJ102326.34+123542.5 & -14.49 & 2739 &    -- &  84 &  $<$0.44 &  $<$18.9 &  $<$0.29 & 32.96 &  -1.9 &  -5 & -- \\
SDSSJ102339.36+123725.6 & -15.16 & 2723 &    -- &  83 &  $<$0.42 &  $<$18.0 & $<$-0.001 & 32.95 &  23.4 &  -5 & -- \\
SDSSJ104140.97+134929.6 & -14.34 & 1271 &    -- &  83 &  $<$0.45 &   $<$5.4 & $<$-0.20 & 31.57 &   1.0 &  -5 & -- \\
SDSSJ104435.28+135622.7 & -12.20 &  633 &    -- &  82 &  $<$0.50 &   $<$0.5 & $<$-0.37 & 28.91 &  -1.7 &  -5 & -- \\
                  LeG14$^a$ & -11.75 &  886 &    -- &  82 &  $<$0.45 &   $<$1.5 &  $<$0.27 & 30.34 & -- &  -2 & -- \\
                  LeG17$^a$ & -14.07 & 1013 &    -- &  83 &  $<$0.33 &   $<$1.9 & $<$-0.54 & 30.95 &  -1.1 &  -2 & -- \\
                NGC3377$^a$ & -19.18 &  684 &    -- & 207 &  $<$0.63 &   $<$1.9 & $<$-2.59 & $^{\star}$30.25 & -- &  -5 & -- \\
                NGC3379$^a$ & -20.03 &  897 &    -- & 207 &  $<$0.59 &   $<$1.7 & $<$-2.98 & $^{\star}$30.21 &  -1.2 &  -5 & -- \\
                NGC3384$^a$ & -19.57 &  733 &    -- &  82 &  $<$0.35 &   $<$1.1 & $<$-2.97 & $^{\star}$30.32 & -- &  -3 & -- \\
            CGCG066-026$^a$ & -13.10 &  541 &    -- &  82 &  $<$0.43 &   $<$0.3 & $<$-1.01 & 28.54 &  -1.4 &  -5 & -- \\
SDSSJ104926.70+121528.0 & -14.43 & 1321 & 1338$\pm$ 12 &  47 &  0.31$\pm$0.08 &   4.1 & -0.35 & 31.87 & 102.6 &  -5 & $c$ \\
                UGC5944 & -15.15 & 1073 &    -- &  83 &  $<$0.35 &   $<$1.0 & $<$-1.25 & $^{\star}$30.22 &  -1.1 &  -5 & -- \\
                NGC3412$^a$ & -18.93 &  853 &    -- & 207 &  $<$0.56 &   $<$1.7 & $<$-2.54 & $^{\star}$30.27 &  -- &  -2 & -- \\
SDSSJ105101.51+132000.5 & -11.97 &  656 &    -- &  82 &  $<$0.37 &   $<$0.4 & $<$-0.40 & 29.12 &  -2.7 &  -5 & -- \\
SDSSJ105131.35+140653.2 & -11.34 &  832 &  816$\pm$ 9 &  32 &  0.22$\pm$0.05 &   0.5 &  0.01 & 29.87 &  -10.6 &  -5 & $c$ \\
SDSSJ105204.79+150149.7 & -12.47 &  828 &    -- &  82 &  $<$0.47 &   $<$1.1 & $<$-0.12 & 29.87 &  -1.3 &  -5 & -- \\
SDSSJ105219.51+110235.6 & -13.45 &  824 &    -- &  82 &  $<$0.43 &   $<$0.9 & $<$-0.60 & 29.75 &  -2.4 &  -5 & -- \\
                UGC6062 & -19.31 & 2629 &    -- & 209 &  $<$0.73 &  $<$30.5 & $<$-1.43 & 32.92 &  -1.2 &  -1 & -- \\
                NGC3489$^a$ & -19.26 &  692 &  692$\pm$ 2 & 107 &  0.86$\pm$0.06 &   2.6 & -2.48 & $^{\star}$30.41 & -- &  -1 & $a$ \\
                  IC676 & -19.42 & 1414 & 1421$\pm$ 2 & 176 &  1.33$\pm$0.07 &  21.9 & -1.62 & 32.12 &   33.6 &  -1 & $a$ \\
SDSSJ111445.02+123851.7$^a$ & -10.86 &  582 &  629$\pm$ 3 &  44 &  0.62$\pm$0.04 &   0.4 &  0.03 & 28.36 &   5.5 &  -5 & $a$ \\
                 IC2684 & -12.42 &  648 &  590$\pm$ 2 &  25 &  0.57$\pm$0.03 &   0.4 & -0.52 & 28.79 &   2.3 &  -5 & $a$ \\
SDSSJ111701.18+043944.2 & -13.40 & 1446 & 1441$\pm$ 24 &  96 &  0.30$\pm$0.07 &   4.8 &  0.13 & 31.90 &   5.2 &  -5 & $c$ \\
SDSSJ112224.02+125846.4 & -11.54 &  626 &    -- &  82 &  $<$0.40 &   $<$0.3 & $<$-0.38 & 28.49 &  -2.7 &  -1 & -- \\
                 IC2782 & -14.44 &  860 &    -- &  82 &  $<$0.44 &   $<$0.8 & $<$-1.08 & 29.69 &  -1.5 &  -5 & -- \\
                 IC2787 & -13.33 &  708 &    -- &  82 &  $<$0.37 &   $<$0.3 & $<$-1.00 & 28.96 &  -2.2 &  -5 & -- \\
                  IC692 & -17.31 & 1157 & 1156$\pm$ 3 &  52 &  3.59$\pm$0.06 &  37.2 & -0.55 & 31.61 &   8.9 &  -5 & $a$ \\
                NGC3773 & -17.31 &  987 &  983$\pm$ 3 &  90 &  3.06$\pm$0.06 &   8.7 & -1.18 & 30.21 &  63.6 &  -2 & $a$ \\
                  IC719 & -17.87 & 1849 & 1848$\pm$18 & 231 &  4.08$\pm$0.10 &  53.9 & -0.61 & $^{\star}$31.87 &  1.1 &  -2 & $a$ \\
                UGC6655 & -13.66 &  748 &  750$\pm$ 3 &  55 &  1.26$\pm$0.07 &   1.0 & -0.66 & 28.81 & 121.7 &   -1 & $a$ \\
 2MASXJ11434609+1342273 & -16.58 & 2920 & 2926$\pm$ 10 &  42 &  0.22$\pm$0.05 &  11.2 & -0.77 & 33.14 & 102.2 &  -2 & $c$ \\
SDSSJ114516.18+135221.2 & -14.56 & 2957 &    -- &  84 &  $<$0.37 &  $<$19.1 &  $<$0.27 & 33.16 &  -1.5 &  -1 & -- \\
 2MASXJ11460404+1134529 & -17.89 & 2977 & 2922$\pm$14 & 168 &  0.72$\pm$0.06 &  37.8 & -0.77 & 33.17 &  75.0 &  -1 & $b$ \\
                NGC4880 & -18.36 & 1377 &    -- & 200 &  $<$0.63 &   $<$2.9 & $<$-2.08 & 30.71 &  -1.6 &  -1 & -- \\
SDSSJ134757.45+041850.6 & -12.01 &  954 &    -- &  83 &  $<$0.49 &   $<$2.4 &  $<$0.38 & 30.79 &  -1.6 &  -5 & -- \\
SDSSJ135142.91+052647.4 & -15.77 & 1241 &    -- &  83 &  $<$0.46 &   $<$5.9 & $<$-0.73 & 31.85 &  -1.1 &  -1 & -- \\
  MAPS-NGPO\_559\_1243538 & -13.83 &  977 &    -- &  83 &  $<$0.43 &   $<$2.3 & $<$-0.36 & 30.91 &  17.6 &  -1 & -- \\
                UGC8799 & -14.42 & 1132 &    -- &  83 &  $<$0.49 &   $<$2.0 & $<$-0.65 & $^{\star}$30.61 &  -1.1 &  -5 & -- \\
                NGC5338 & -17.17 &  804 &  803$\pm$ 3 & 135 &  0.57$\pm$0.06 &   2.2 & -1.72 & $^{\star}$30.54 &  24.9 &  -2 & $a$ \\
SDSSJ135502.70+050525.2 & -15.13 & 1396 &    -- &  83 &  $<$0.49 &   $<$7.6 & $<$-0.36 & 32.05 &   -1.4 &  -5 & -- \\
SDSSJ135621.31+051944.2 & -14.54 & 1395 &    -- &  83 &  $<$0.45 &   $<$7.0 & $<$-0.16 & 32.05 &  -1.2 &  -5 & -- \\
            CGCG046-013 & -16.41 & 1516 &    -- &  83 &  $<$0.50 &   $<$8.6 & $<$-0.82 & 32.17 &  -1.5 &  -5 & -- \\
SDSSJ135723.57+053425.2 & -15.01 & 1055 &    -- &  83 &  $<$0.45 &   $<$3.7 & $<$-0.62 & 31.36 &   0.6 &  -5 & -- \\
                UGC8986 & -17.78 & 1232 &    -- & 207 &  $<$0.70 &   $<$8.5 & $<$-1.37 & 31.78 &  -1.2 &   -2 & -- \\
SDSSJ142043.54+040837.0 & -15.03 & 1704 &    -- &  83 &  $<$0.43 &   $<$8.6 &  $<$0.27 & 32.34 &  3.5 &  -2 & -- \\
SDSSJ144329.18+043153.4 & -15.81 & 1745 & 1716$\pm$ 2 &  74 &  1.36$\pm$0.06 &  28.6 & -0.06 & 32.38 &  13.5 &  -5 & $a$ \\
                NGC5770 & -18.41 & 1454 &    -- & 200 &  $<$0.63 &   $<$5.4 & $<$-1.83 & $^{\star}$31.39 & -- &  -2 & -- \\
SDSSJ152655.37+094657.5 & -13.96 & 1858 &    -- &  83 &  $<$0.41 &   $<$7.7 &  $<$0.11 & 32.26 &  0.2 &  -5 & -- \\
                 IC1131 & -18.05 & 2017 &    -- & 200 &  $<$0.63 &  $<$18.0 & $<$-1.16 & 32.71 & -- &  -5 & -- \\
                NGC6014 & -20.04 & 2491 &    -- & 200 &  $<$0.63 &  $<$25.4 & $<$-1.80 & 33.08 & -- &  -2 & -- \\
\hline \hline
\end{tabular}
\footnotetext[1]{Leo group member}
\end{minipage}
\end{center}
\end{table*}

In the monolithic scenario \citep{1974MNRAS.166..585L, 2002MNRAS.335..335C},  early-type galaxies (ETGs) form in a single, short, and highly efficient
burst of star formation at high redshift ($z > 3$) followed by a more quiescent evolution until today as the residual gas flows into the nuclear region of the galaxy. In this model, the bulk of the stellar population was formed at early epochs and ETGs appear today as gas-poor systems dominated by an old stellar population.
On the other hand, hierarchical structure formation scenarios suggest that ETGs seen today in clusters have followed different evolutionary paths compared to the galaxies in LDEs \citep{1999AJ....117.1651G},
which are still undergoing hierarchical assembly and therefore are expected to be on average younger with more heterogenous properties 
\citep{1992AJ....104.1039S, 1997ASPC..116..469B}.
\citet{1980ApJ...237..692L} first pointed out that field ellipticals may have experienced a more extended star formation history with respect to their cluster counterparts. \citet{1994AJ....108.2054R} found an intermediate-age population in E/S0 galaxies in LDEs, suggesting a wider spread in the epochs of major star formation activity.
Spectroscopic observations of 26 ETGs located in voids \citep{2008AJ....136....1W} indicate that even though there is a range of old ($>$ 9 Gyr) and young ($<$ 5 Gyr) systems, there appears to be a greater proportion of young galaxies in voids than in clusters.

The finding of gaseous structures, 
  such as rings and discs, mostly in ETG field galaxies, can give further indication that  merging events with gas-rich objects are taking place today in LDEs
\citep[see review by][]{1990ApJ...364L..33S}.
Detailed studies of nearby ETGs have shown that a fraction of them contains neutral hydrogen \citep{1980ApJ...242..931S, 1985AJ.....90..454K, 1994A&A...286..389H} and they
can host extended \hi structures around them \citep{2007A&A...465..787O}, accompanied in
some cases by traces of ongoing or recent star formation \citep{2000AJ....119.1180S, 2007NewAR..51....3S, 2008MNRAS.385.1965H}.

The presence of neutral hydrogen in ETGs may be explained as the consequence  of a recent accretion of a gas-rich
satellite or of a merging process between similar-size galaxies \citep{1985AJ.....90..454K, 1991A&A...243...71V, 2002MNRAS.333..481B}. Simulations show that spheroidal galaxies with extended gaseous discs can form in spiral mergers \citep{2002MNRAS.333..481B, 2004cbhg.symp..421B}, as the result of the subsequent infall of the \hi from the tidal tails to the main merging remnant. This process can rebuild the gaseous disc on the timescale of few Gyrs leading to an elliptical or a S0 galaxy with an extended gaseous component \citep{1995AJ....110..140H}.

An alternative scenario foresees that the gas is accreted
from the intergalactic medium via a 'cold mode', where the gas cools along filamentary structures  without being
shock-heated to the virial temperature of the halo  \citep[T $\sim 10^5$ K;][]{2005MNRAS.363....2K, 2006MNRAS.366.1529M}.

When the \hi distribution is mapped at high resolution, gas-rich ETGs
show regular \hi structures appearing as low column density ($< 5 \times 10^{20}$ \cmsq) discs or rings in rotation \citep{1991A&A...243...71V, 2007A&A...465..787O}. The regular kinematics of these structures suggests that they are relatively old and long-lived (a few Gyrs). \hi tails or  clouds offset from the optical counterpart are also found in some systems \citep{2006MNRAS.371..157M, 2007A&A...465..787O}  which is indicative of recent interactions or of a gas accretion process.

Nonetheless, the rate of detection of neutral hydrogen in nearby ETGs varies broadly depending on the depth of observations and the choice of the sample \citep{2002ASPC..273..215S, 2006MNRAS.371..157M, 2007A&A...474..851D}.
It is well known that the \hi content of late-type galaxies in dense cluster decreases when compared to systems of the same morphological type in lower density regions \citep{1983ApJ...267...35G, 1986ApJ...306..466H, 2001ApJ...548...97S}. But how does the \hi content of ETGs vary with the environment?
Variations in the detection rate of \hi in ETGs have already been noticed in previous studies  \citep{1980ApJ...240L..87H, 1983AJ.....88.1719G, 1994A&A...286..389H, 2007A&A...465..787O}, confirming that the probability of finding ETGs with gas is higher in lower density environments.
The \hi Parkes All-Sky Survey \citep[HIPASS;][]{2004MNRAS.350.1195M} in the southern hemisphere provides a large database at 21-cm which has been recently used to
study the \hi content of 2500 nearby ETGs (E and S0) extracted from the RC3 catalog \citep{1991trcb.book.....D}. HIPASS  detected \hi with a rate of 6\% for elliptical and 13\% for S0 galaxies \citep{2002ASPC..273..215S}. However this preliminary analysis of the HIPASS data is plagued by confusion, since 30\%-50\% of the \hi detected ellipticals have more than one neighbor with similar optical velocity within the 15$^{\prime}$ Parkes beam, and no distinction between different environments has been  analysed or discussed. These results suggest the need for a systematic attempt to determine the \hi properties of ETGs with a better spatial resolution, extending this analysis to a more uniform  sample of galaxies in different environments.

\begin{figure*}
\includegraphics[width=19cm]{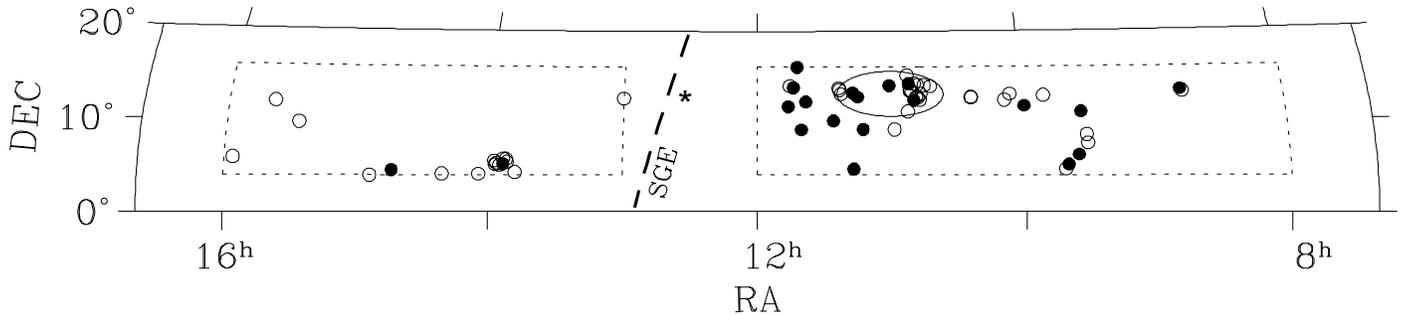} 
\caption{The sky distribution of the selected sample of ETGs in Right Ascension and Declination.
Filled dots correspond to galaxies detected at 21-cm, while empty dots show gas-poor ETGs.  The ellipse indicates the area of the Leo group. The star between the two regions where we have defined our sample gives the position of M87, while the thick dashed line traces to the super galactic plane.}
\label{sky}
\end{figure*}

In Paper I we have used data from the ALFALFA survey \citep{2005AJ....130.2598G, 2007AJ....133.2569G}, an
ongoing blind survey  at 21-cm performed with the Arecibo telescope,
to start a more uniform and systematic study of the \hi content of ETGs, by analysing it in the Virgo cluster.
The advantage of using ALFALFA with respect to a previous blind \hi survey such as HIPASS is twofold: its higher sensitivity and the smaller beam size which reduces the possibility of confusing the \hi detections.
We have defined an optical sample of 457 ETG, extracted from the Virgo Cluster Catalog \citep[VCC][]{1985AJ.....90.1681B}, which are brighter
than the VCC completeness limit at B$_T$ = 18.0. We have correlated this optical sample with the catalog of detected \hi sources from
ALFALFA. Only 9 out of 457 ETGs (roughly 2\%) are detected in \hi,
with the majority of ETGs with gas having peculiar morphologies and being located in the outer regions of the cluster.

Here we extend our analysis to a low density sample of ETGs, to compare the \hi detection rate 
in different environments.

The paper is organised as follows: in Section 2 we describe the sample selection, in Section 3 we give an overview of the results, presenting the 21-cm detections, and their optical properties.
In Section 4 we discuss the role of the environment on the \hi detection rate of ETGs, and the evidence of recent star forming activity in the sample, and in
Section 7 we present our  conclusions.

\section{Sample selection}

The strategy we have followed in this study is
the same as in Paper I for the Virgo cluster.
We define  a--priori an optical sample of  ETGs within a definite radial velocity limit,
which we then search for \hi, using the ALFALFA survey.

The selection of a LDE optical sample of  ETGs  as similar as possible to the Virgo one, in terms of luminosity distribution, and possibly complete to a faint limiting magnitude is the crucial point of this work.
For this aim we have used the Sloan Digital Sky Survey \citep[SDSS;][]{2000AJ....120.1579Y, 2002AJ....123..485S} which provides a homogeneous data set of galaxies with both photometric and spectroscopic measurements in different environments: from voids to groups and rich clusters. Spectra are taken only for a subset of objects,
the so called main spectroscopic sample, consisting of galaxies with $r$-band 
magnitudes $r < 17.77$ and $r$-band 
half-light surface brightnesses $\mu_{50} \leq 24.5$ mag arcsec$^{-2}$ \citep{2002AJ....124.1810S}.

First we selected all the objects in the seventh data release \citep[DR7;][]{2008arXiv0812.0649A} within the sky region where ALFALFA data are currently available with the same cut in radial velocity as in Paper I, i.e.

\begin{itemize}
\item V $<$ 3000 \kms;
\item 8$^h <$ RA $<$ 16$^h$, avoiding the region of the Virgo cluster between 12$^h$ and  13$^h$;
\item $4^{\circ} <$ DEC $< 16 ^{\circ}$.
\end{itemize}

The criterium on the radial velocity was imposed to reach a comparable \hi mass detection limit to the one of Paper I. After this preliminary selection we obtained 307 galaxies of all morphological types.

\citet{2005AJ....129...61B} defined a set of criteria to separate early and late-type galaxies in the SDSS database
 from the value of two outputs by the SDSS pipeline, the spectroscopic parameter $eclass$ and the photometric parameter $fracDev_r$\footnote{$eclass$ is a spectroscopic
parameter giving the spectral type from a principal component
analysis, while $fracDev_r$ measures the fraction of light profile that is fitted by a de Vaucoulers law}: early-type morphologies are defined
by choosing $eclass < 0$ and $fracDev_r > 0.8$. Applying both these cuts we obtained only 4 galaxies. However we realised that this method was not appropriate for our aims. The use of $fracDev_r$ would prevent  to select early-type dwarfs, because their light profiles are generally fitted by an exponential law \citep{1994A&ARv...6...67F}, while the choice of $eclass < 0$ would have automatically excluded
ETGs with emission line spectra, a possibility that cannot be rejected a-priori, especially for our interest in gas-rich ETGs.
Moreover, because the aperture of a SDSS spectroscopic fiber (3") samples only the inner parts of nearby galaxies, we found that, using this criterium on its own, the bulges of some nearby late-type galaxies were mistakenly selected as early-type objects.
Therefore we examined by visual inspection the SDSS multi-colour images of the 307 galaxies and classified 50 of them as early-type (E and S0). 

The completeness of the main spectroscopic sample depends on the magnitude \citep{2002AJ....124.1810S}. The main source of incompleteness is due to blending with saturated stars, an effect which is more significant for brighter and larger galaxies. According to \citet{2002AJ....124.1810S}, at bright magnitudes ($r < 15$) only 5\% of the galaxies in the Zwicky catalog are missed, while from comparison with visual inspections of all objects brigther than $r = 18$ over 22 deg$^2$ of sky, the completeness of the sample is above 99\%.

To remedy the incompleteness of the SDSS spectroscopic sample at bright magnitudes, we have looked for ETGs with V $<$ 3000 \kms in the RC3 catalog \citep{1991trcb.book.....D} and in the Nasa/IPAC Extragalactic Database (NED), and we found that some galaxies  had been missed by our selection from the SDSS database. From the RC3 we added  10 bright ETGs, probably excluded by the SDSS spectroscopic target selection algorithm because blended with saturated stars. From NED we found also one faint dwarf (LeG 14) whose $r$-band magnitude is fainter than the limit of the main sample ($r < 17.77$), and Leo I, which was not included in the SDSS spectroscopic sample because it is resolved.

Thus we have built an apparent magnitude limited and bound in redshift sample of ETGs in LDEs which is complete down to the $r$-band magnitude $r < 17.77$ \citep{2002AJ....124.1810S}. Our final list of targets is composed of 62 objects,  34 ellipticals and 28 S0s. For the galaxies in common with the RC3 catalog \citep{1991trcb.book.....D} we have adopted the RC3  morphological classification.
39 objects have absolute magnitudes M$_B > -17$, and they can be considered as dwarf ETGs.

Figure \ref{sky} shows the sky distribution of the galaxy sample and the regions we have selected, which also include the Leo group; 
in Table 1 the main observational parameters of our targets are displayed.
The columns are as follows:

{\em Column (1).} The name of the galaxy.

{\em Column (2).} The absolute B magnitude of the galaxy. B apparent total magnitudes have been derived from the RC3 when available, otherwise they have been calculated from the $model$\footnote{The model magnitude ($modelmag$) in the SDSS database is defined as the total magnitude calculated by using the de Vaucouleurs or exponential model that best fits the galaxy profile in the $r$ band.} $r$, $g$ magnitudes derived from the SDSS database and converted to the Johnson system using the following relation B=$g$ + 0.17 + 0.47$\times (g-r)$   \citep{2002AJ....123.2121S}. A correction to the $g$ magnitudes, $\Delta g = 1-$exp[-0.11($g-17$)] , has been applied to compensate for the overestimate of the local sky flux
near bright large galaxies performed by the SDSS photometric pipeline \citep{2005MNRAS.361.1287M}.

{\em Column (3).} The optical heliocentric velocity of the galaxy extracted from  NED in  \kms.

{\em Column (4).} Heliocentric velocity of the \hi detections, measured as the midpoint of the 50\% level of the peak flux density. 

{\em Column (5).} Velocity width of the \hi line profile, measured at half peak value. More details on the way it has been estimated can be found in \citet{2007AJ....133.2569G}.

{\em Column (6).} Integrated flux density $F$ of the source in Jy \kms and its estimated uncertainty. For non detections, upper limits on the flux density have been derived by measuring the rms of the integrated spectrum, obtained  over a  region of 5$^{\prime} \times$ 5$^{\prime}$ around the optical position. The flux value has then been calculated for a S/N = 5 and a velocity width of 200 \kms if M$_B < -17$ or 80 \kms if M$_B > -17$.

{\em Cols (7).} The \hi mass (and upper limits for non detections) in units of $10^7$ \msun, obtained using the standard equation $M_{HI} = 2.35 \times 10^5$ d$^2_{Mpc}$ $F$ where d$_{Mpc}$ is the assumed distance in Mpc.

{\em Cols (8).} Logarithm in base 10 of the \hi mass-to-stellar light ratio in solar units (and upper limits for non detections).

{\em Cols (9).} Distance modulus. For 12 objects (labeled with a star) direct distance measurements are available from various indicators such as the tip of the red giant branch, surface brightness fluctuations, supernovae, planetary nebulae,  and they have been taken from \citet{2008ApJ...676..184T} and NED\footnote{http://nedwww.ipac.caltech.edu/level5/NED1D/}. For the other galaxies we have adopted the Hubble flow distance moduli calculated in NED, based on the local velocity field given in \citet{2000ApJ...528..655M}, assuming $H_0 = 73$ \kms Mpc$^{-1}.$ The corresponding distances range between 230 kpc (Leo I) and 47 Mpc (2MASX11460404+1134529).

{\em Cols (10).} H$\alpha$ equivalent width in  Angstrom, derived from the SDSS spectra. 
The H$\alpha$ equivalent width of NGC3773 has been taken from Gavazzi et al. (in preparation), because it is not included in the SDSS spectroscopic sample.

{\em Cols (11).} Morphological type of the galaxies in the sample. T=-5 refers to ellipticals, and  T=-3,-2,-1 correspond to S0 galaxies.

{\em Cols (12).} The code of the \hi detections. Given that the signal threshold 
in the ALFALFA catalog is S/N $>$ 6.5 for $\Delta$V $<$ 400 \kms \citep{2007AJ....133.2087S}, we define with
code $a$ firm detections with S/N above the adopted ALFALFA threshold, code $b$ denotes reliable detections with a lower signal-to-noise ratio ($5 <$ S/N $<6.5$) which have an optical counterpart with a similar redshift, while code $c$ refers to possible detections which need to be corroborated with deeper observations.

\begin{figure*}
\begin{center}
\includegraphics[width=4.5cm]{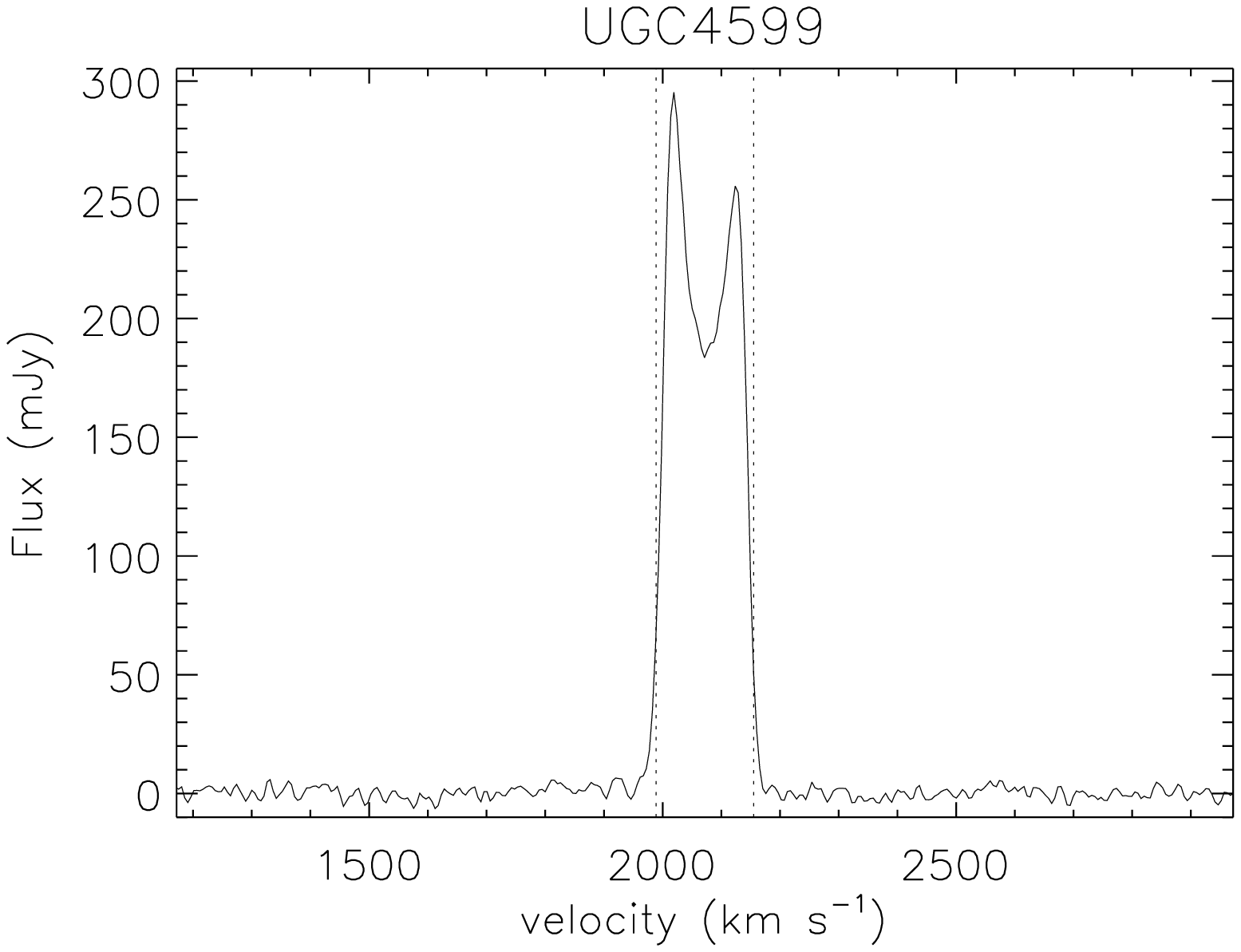} 
\includegraphics[width=4.5cm]{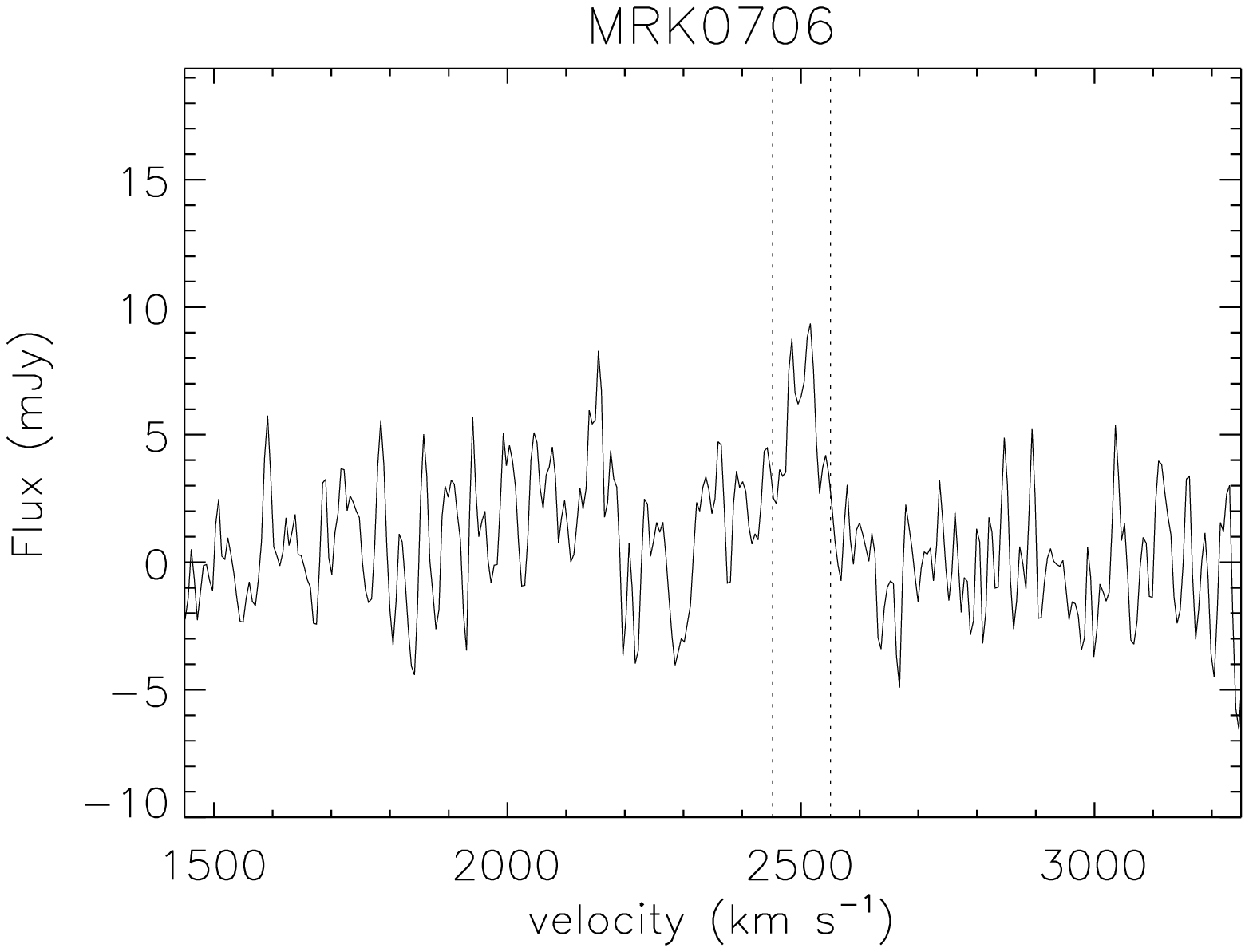} 
\includegraphics[width=4.5cm]{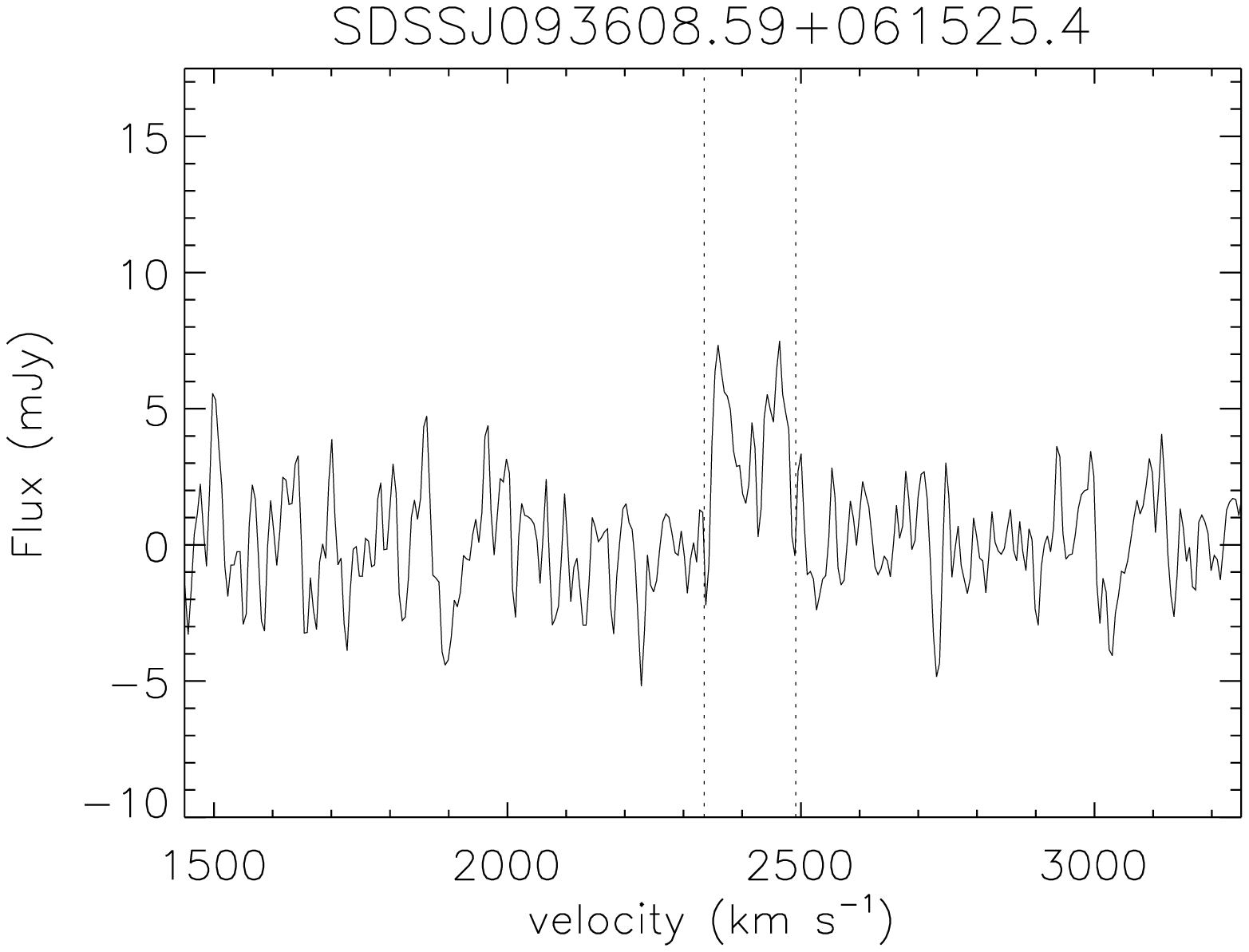} 
\includegraphics[width=4.5cm]{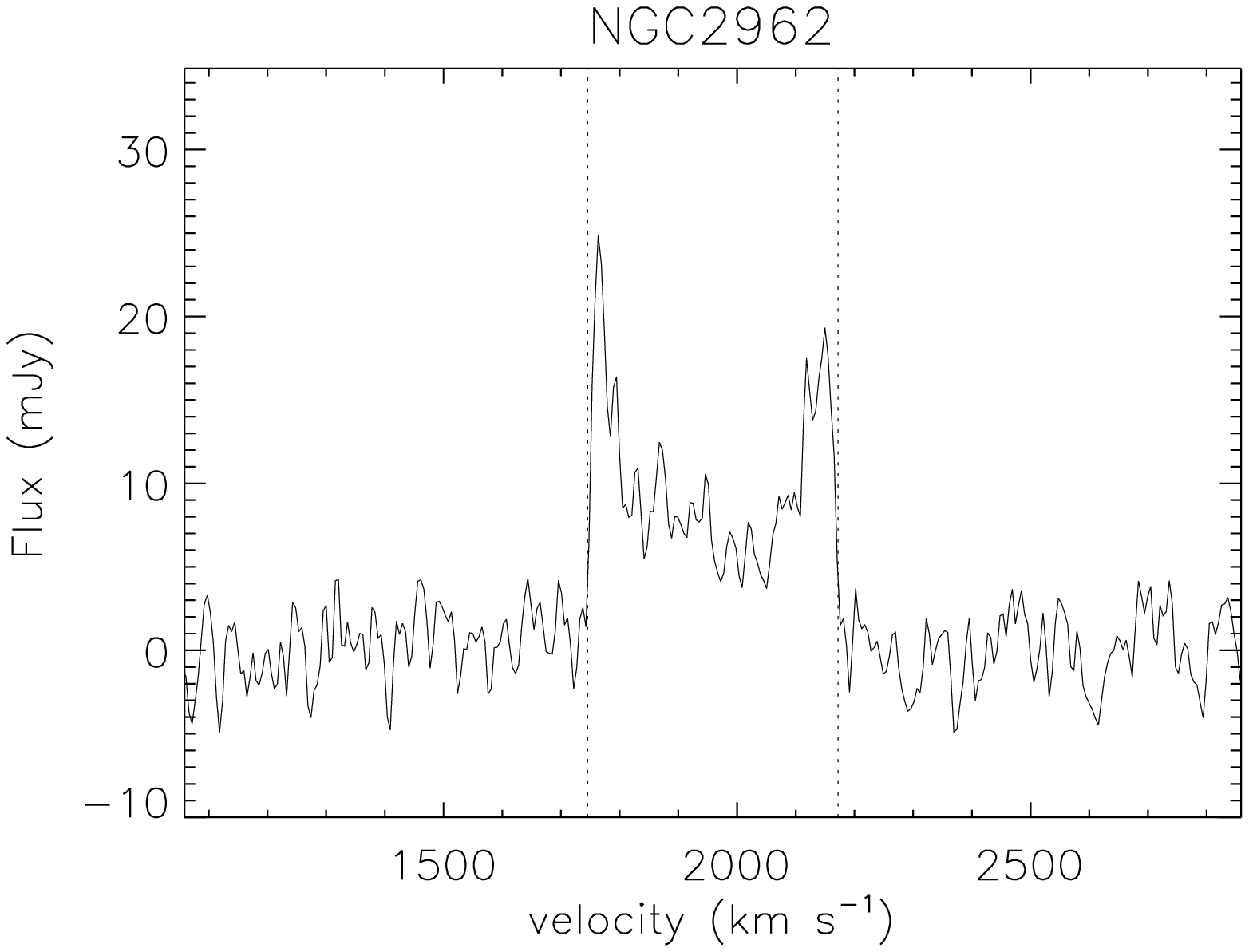} 
\includegraphics[width=4.5cm]{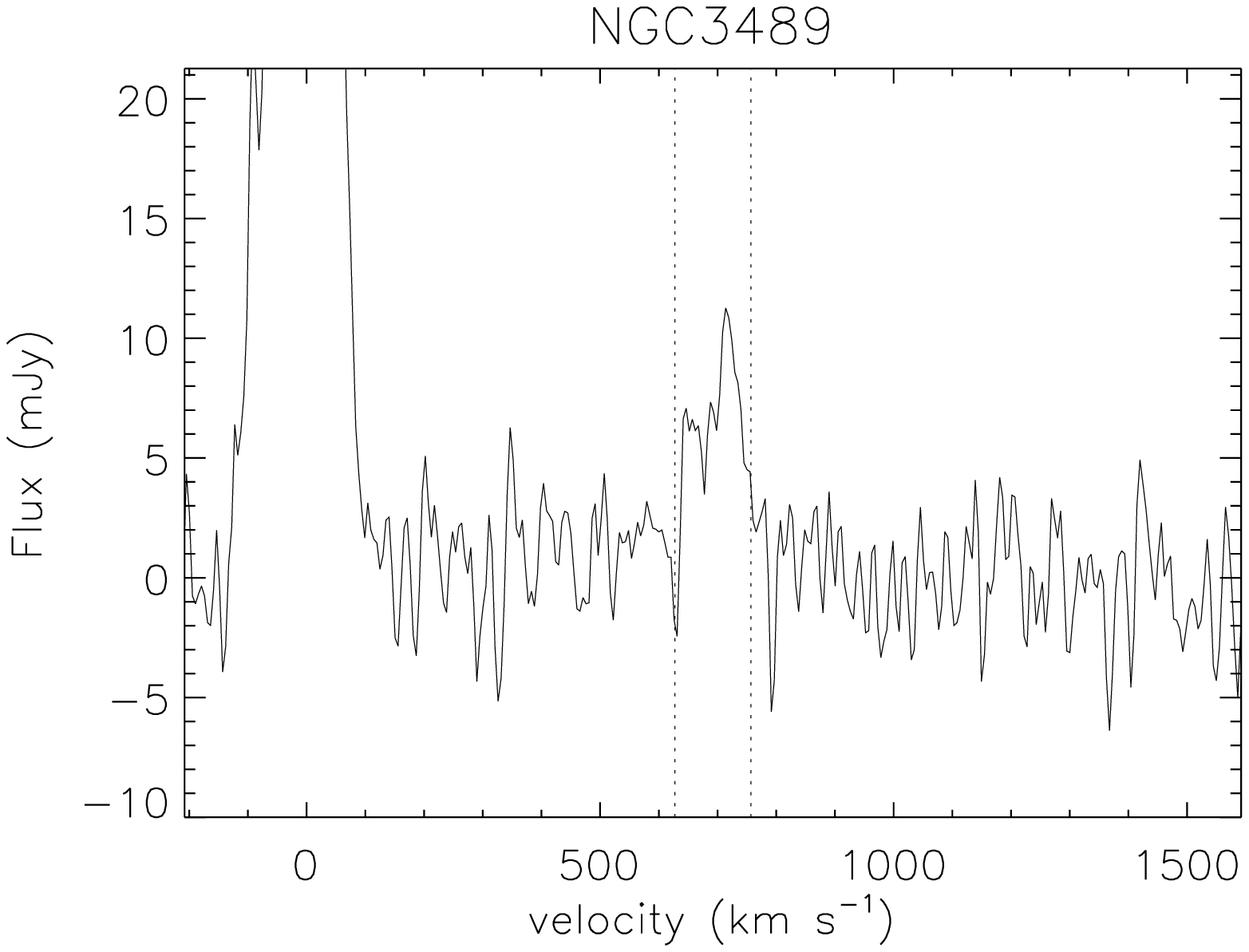} 
\includegraphics[width=4.5cm]{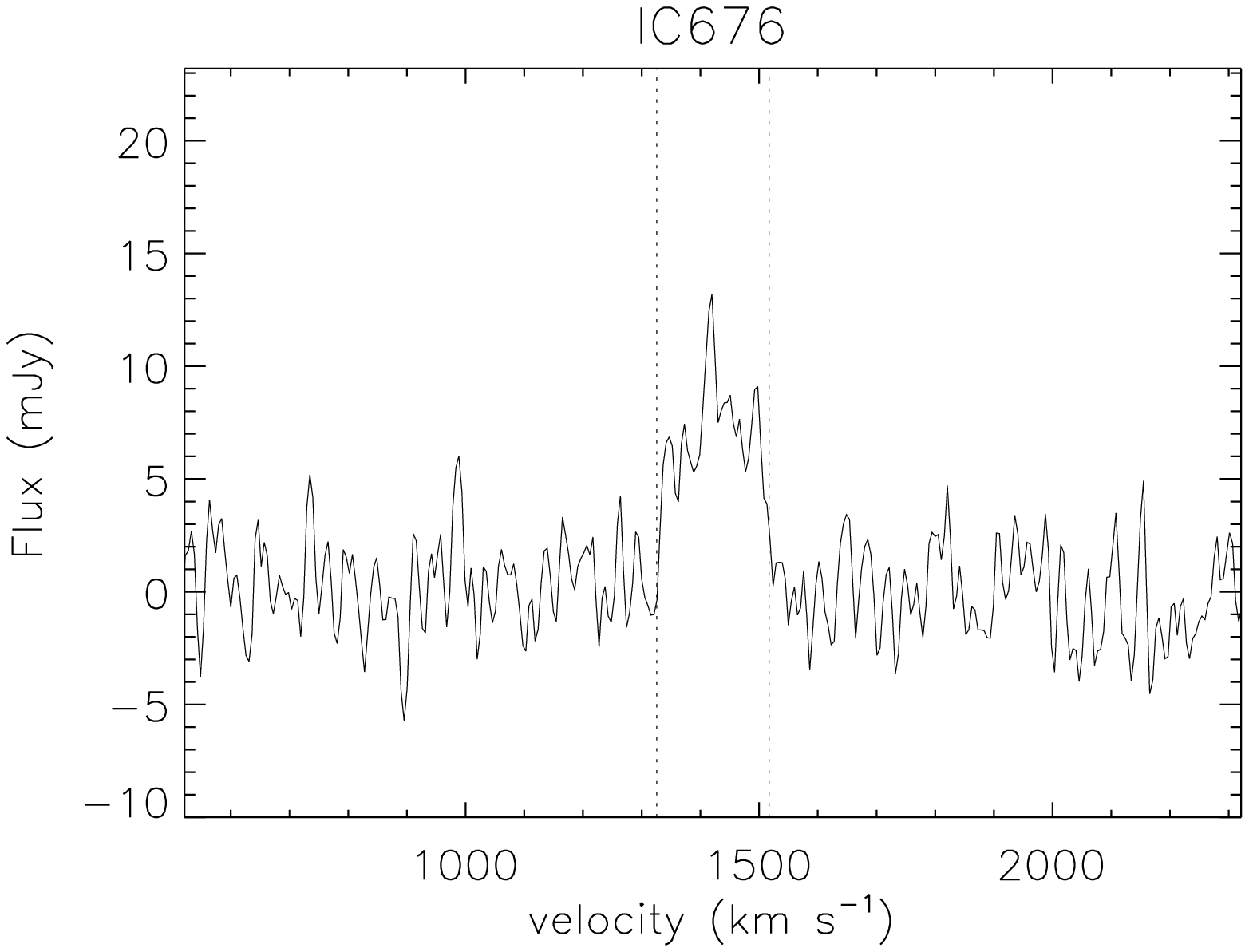} 
\includegraphics[width=4.5cm]{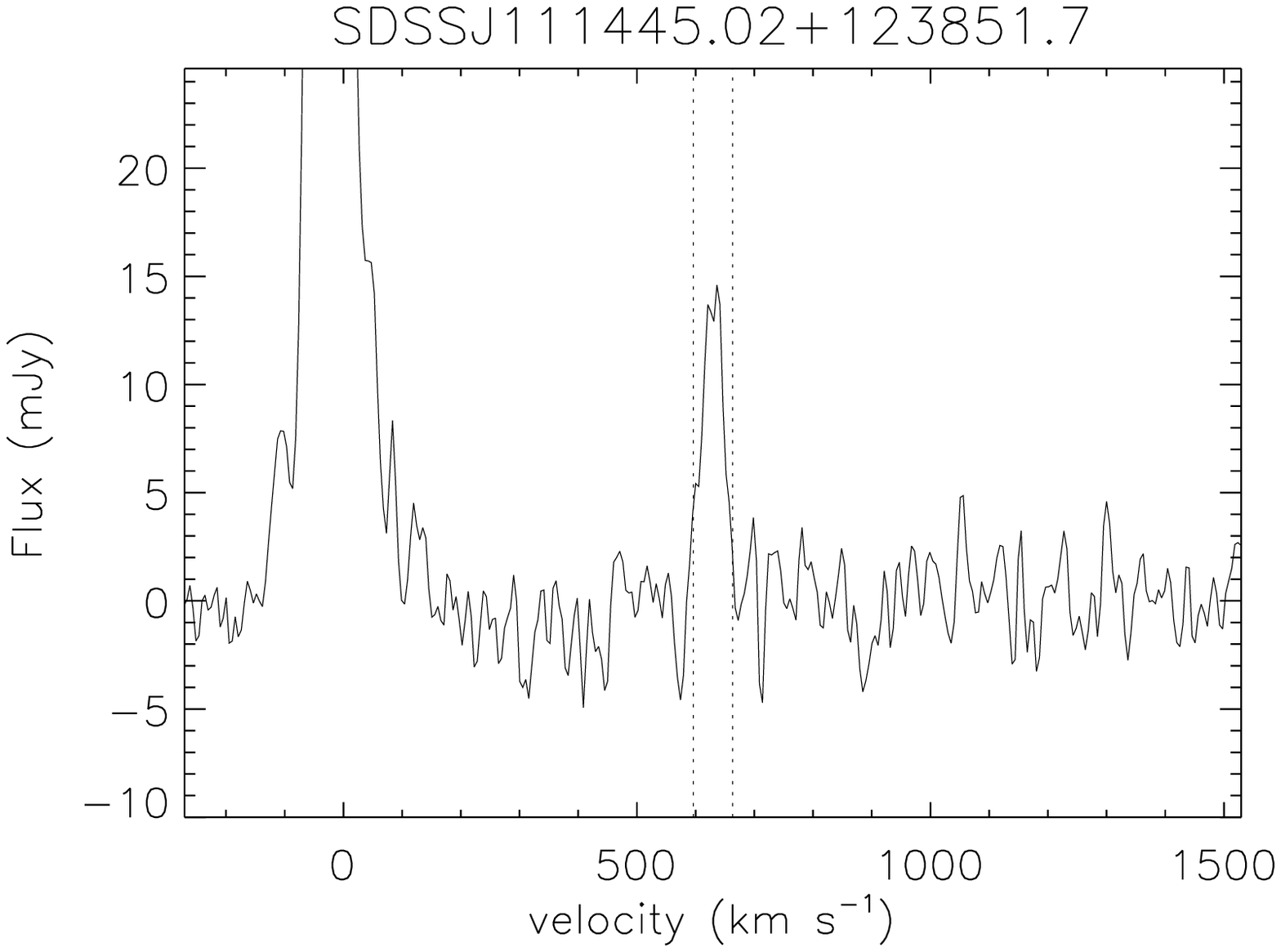} 
\includegraphics[width=4.5cm]{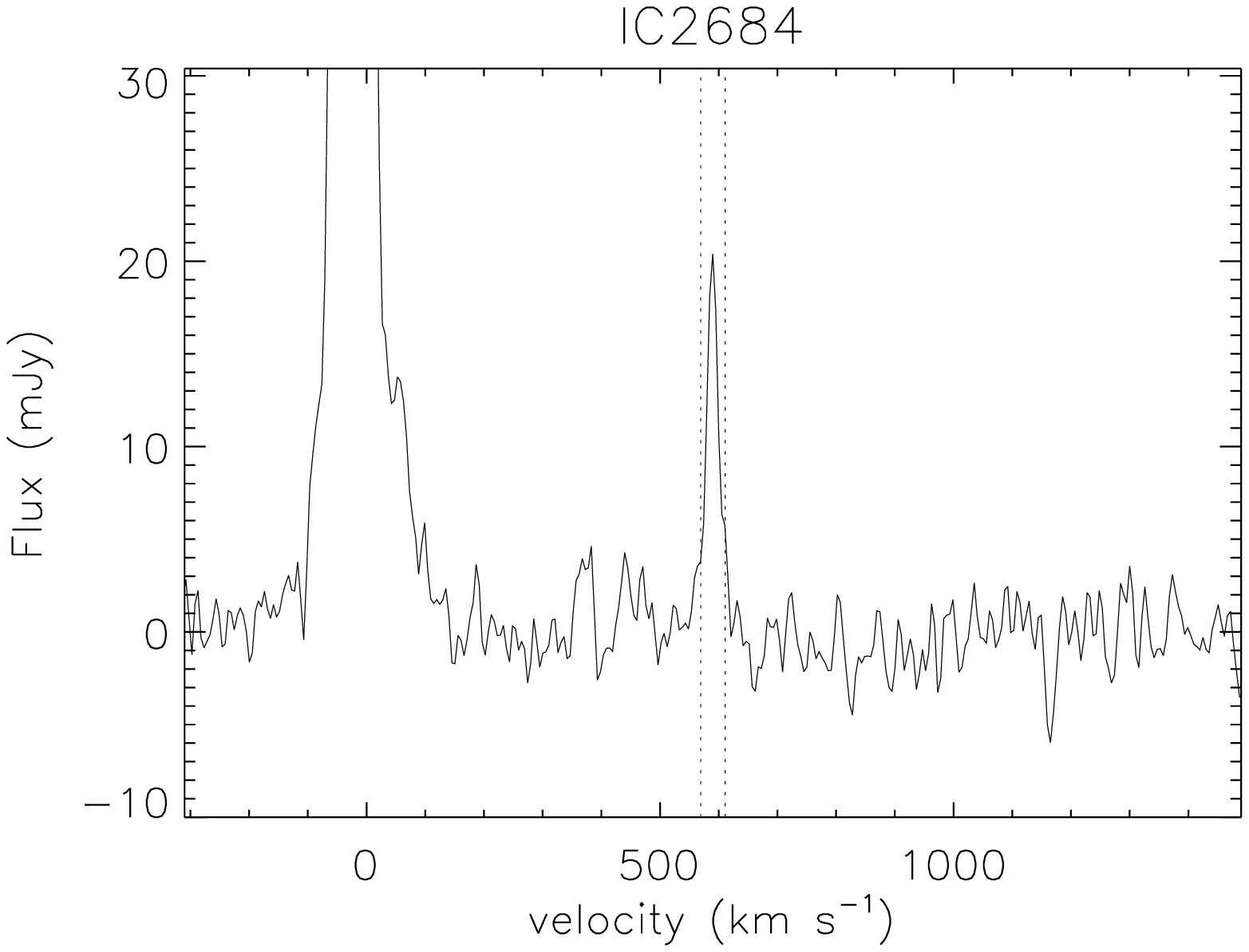} 
\includegraphics[width=4.5cm]{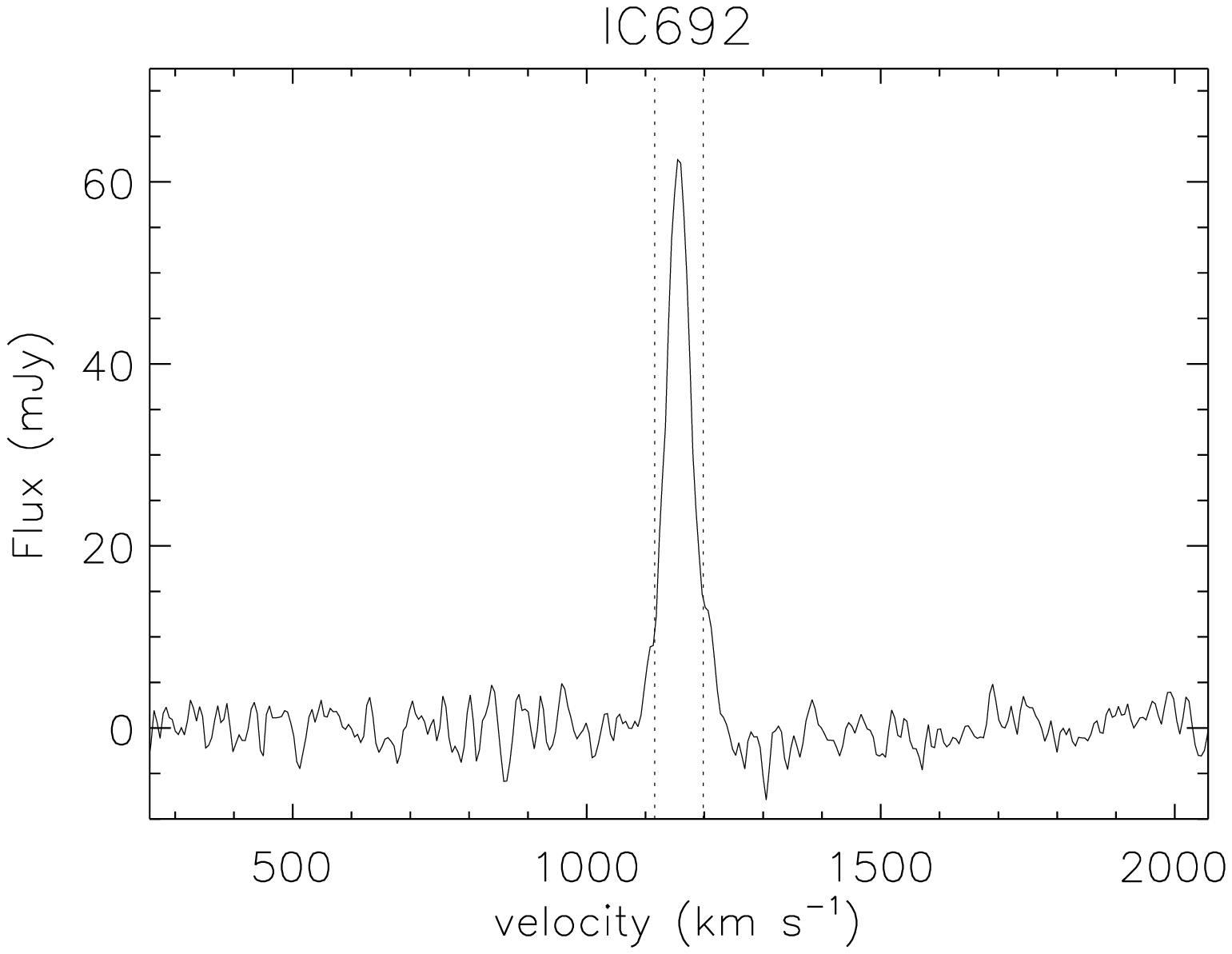} 
\includegraphics[width=4.5cm]{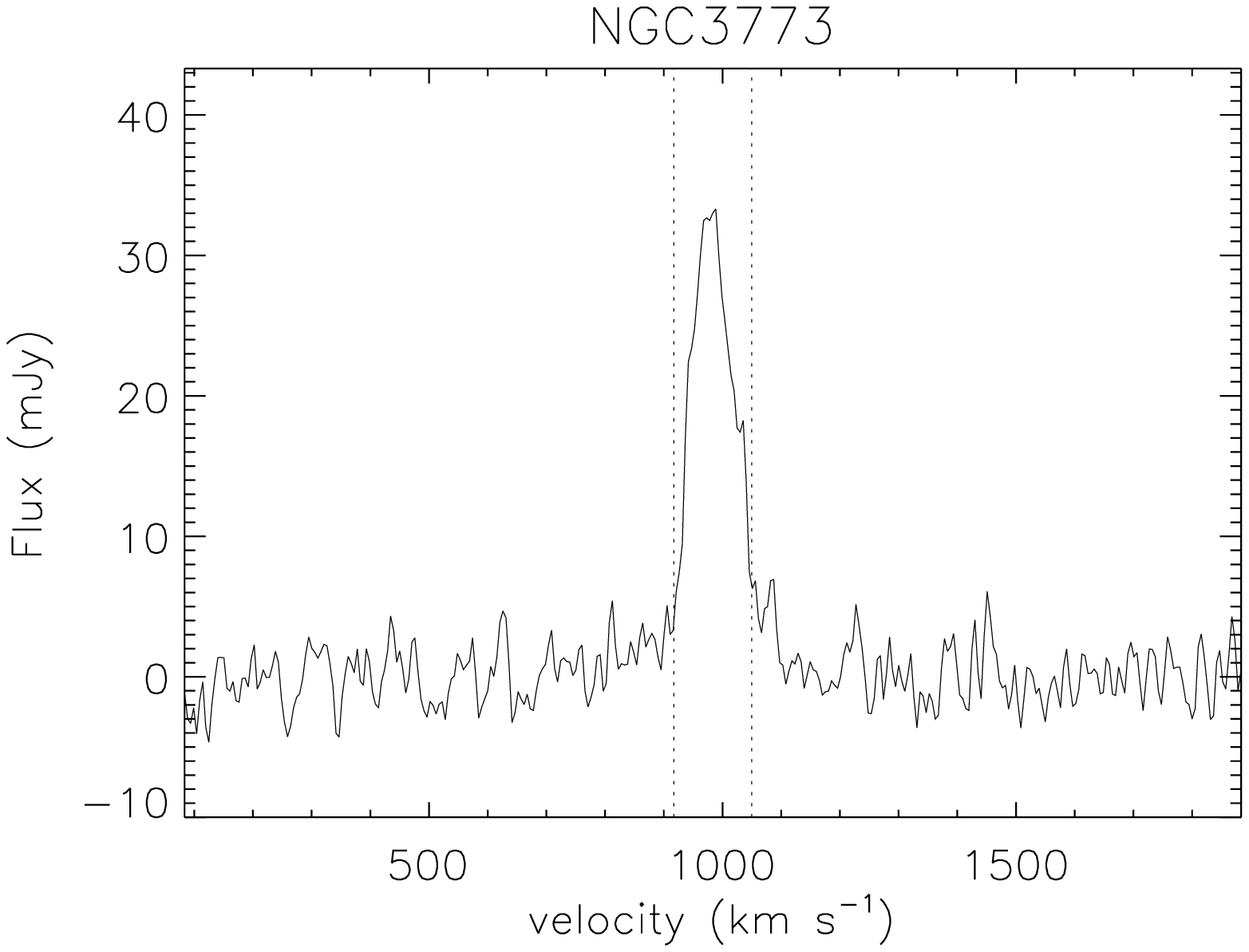} 
\includegraphics[width=4.5cm]{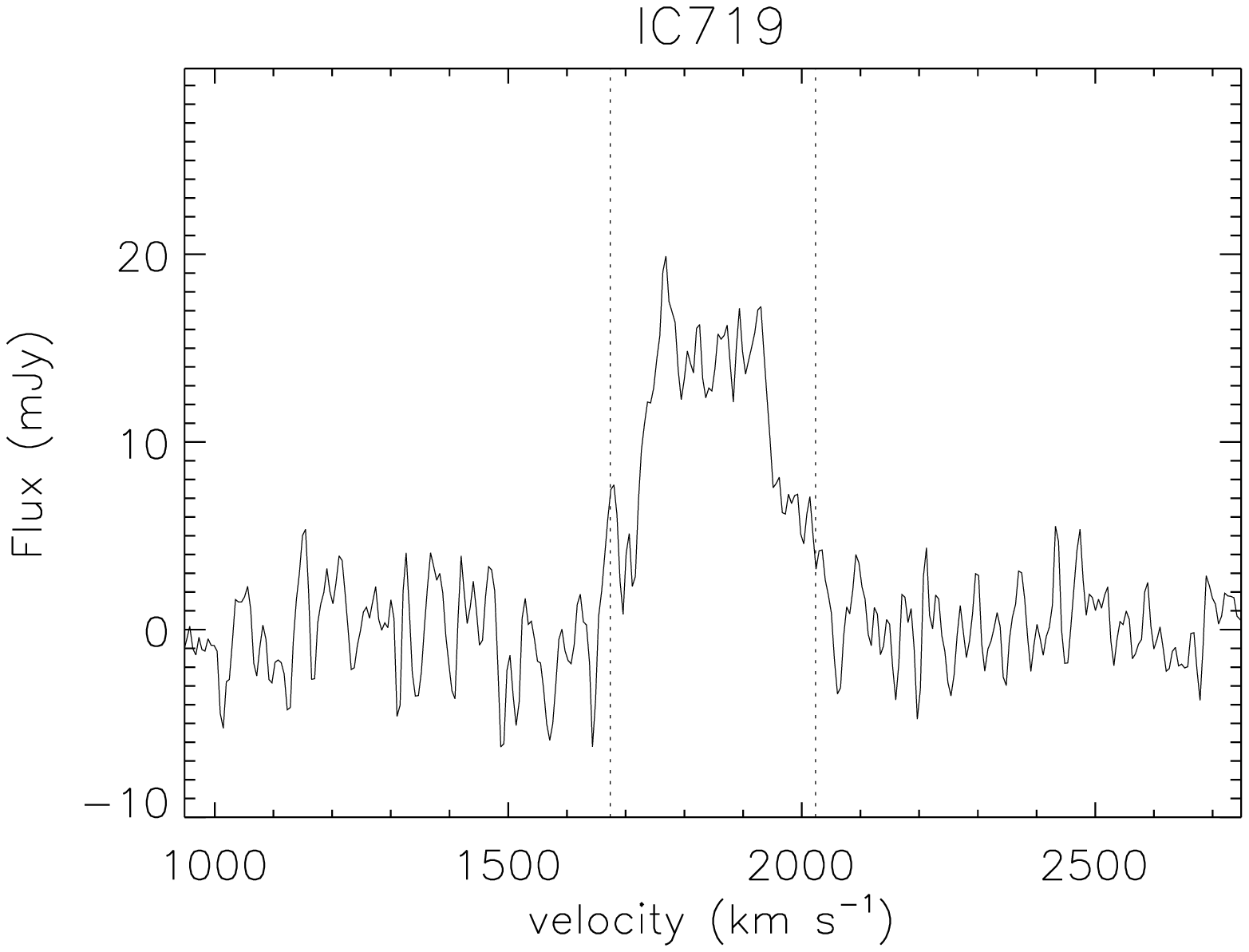} 
\includegraphics[width=4.5cm]{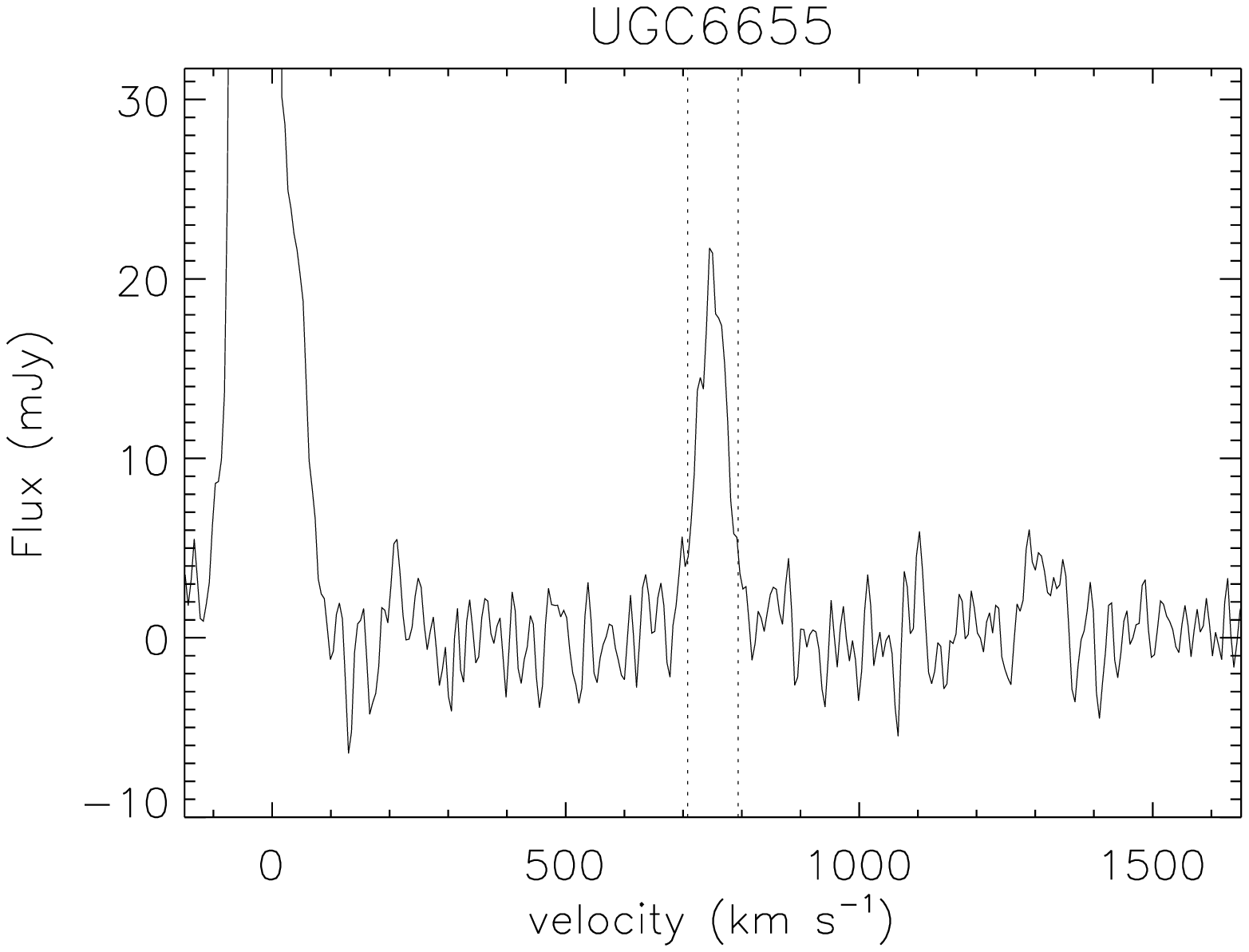} 
\includegraphics[width=4.5cm]{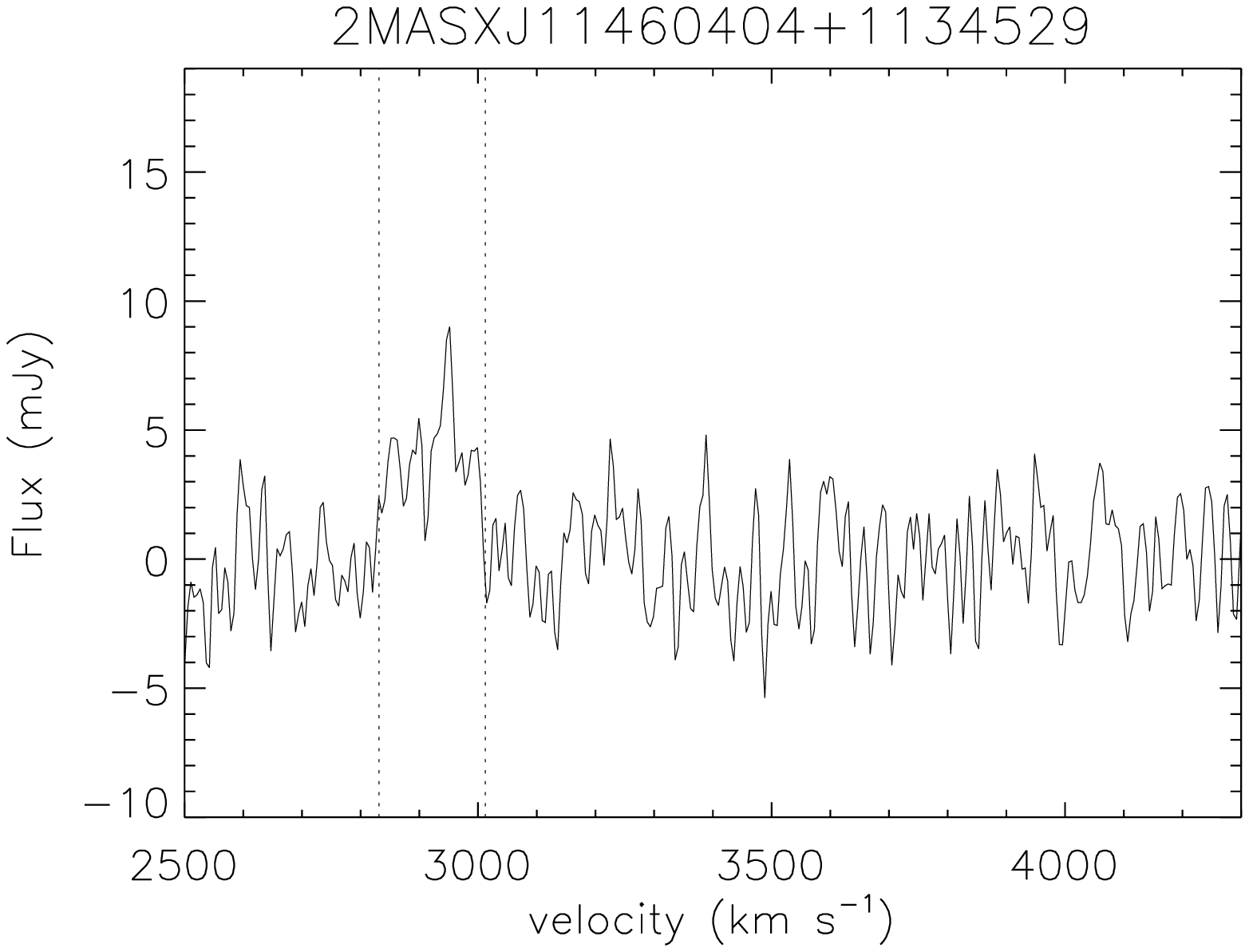} 
\includegraphics[width=4.5cm]{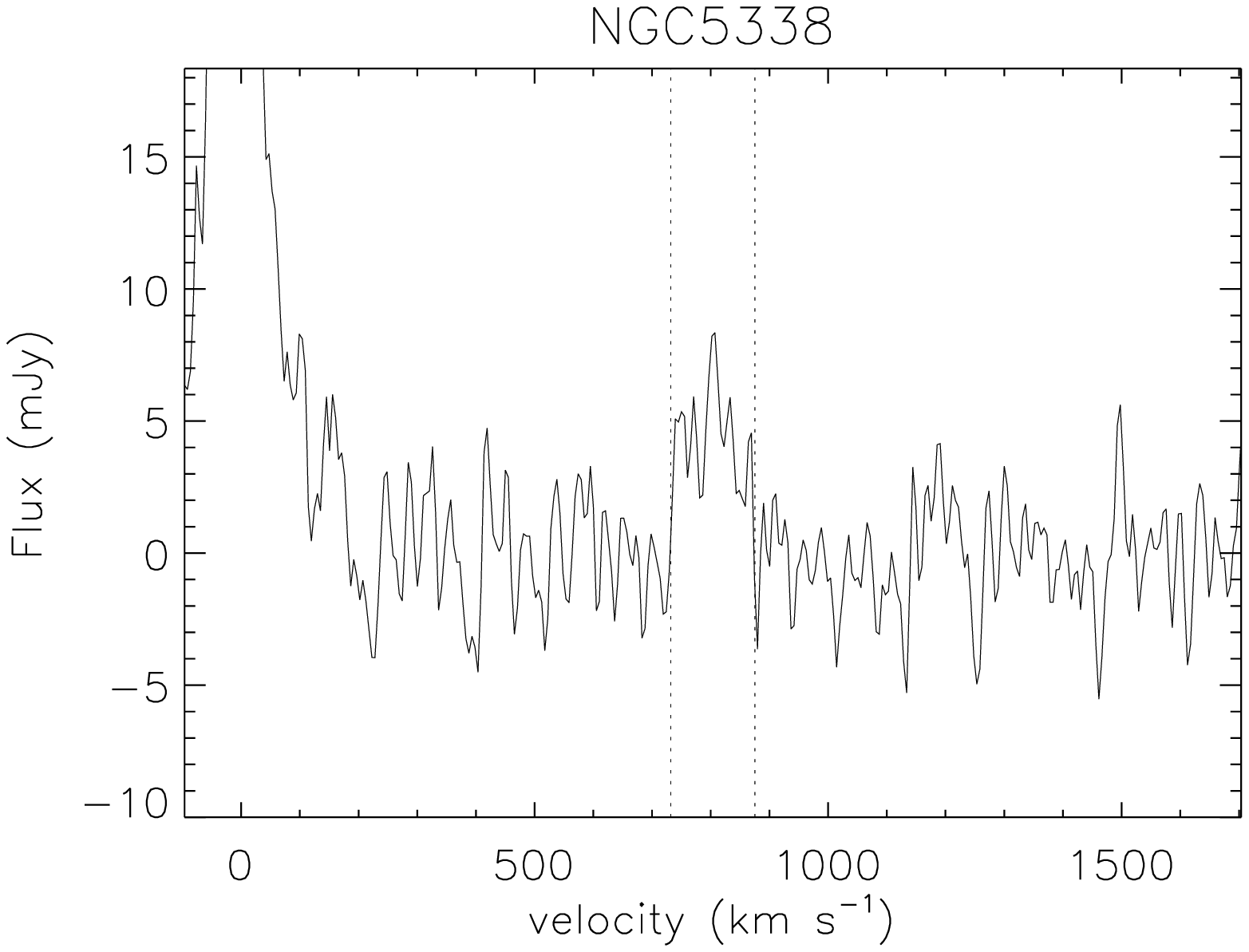} 
\includegraphics[width=4.5cm]{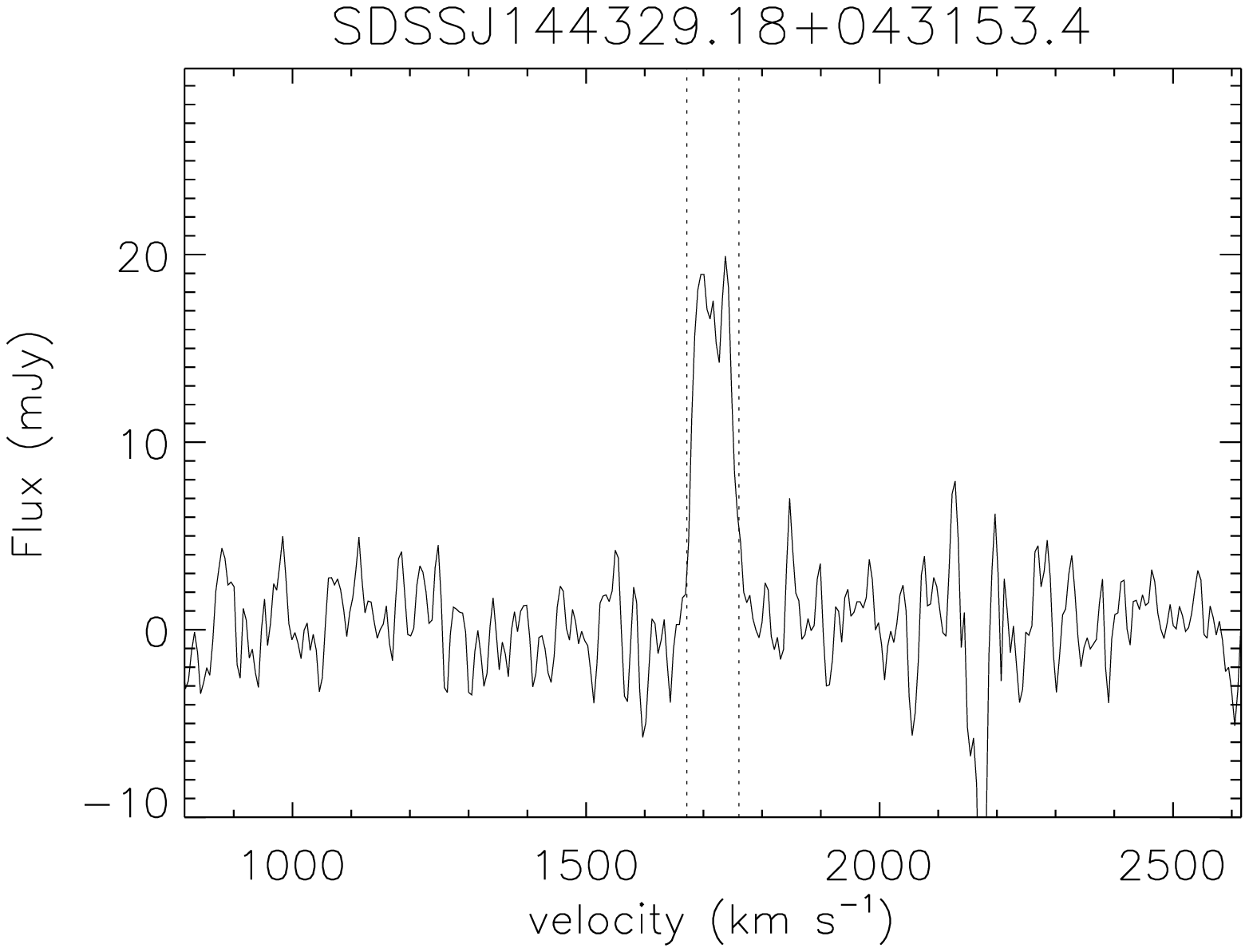} 
\includegraphics[width=4.5cm]{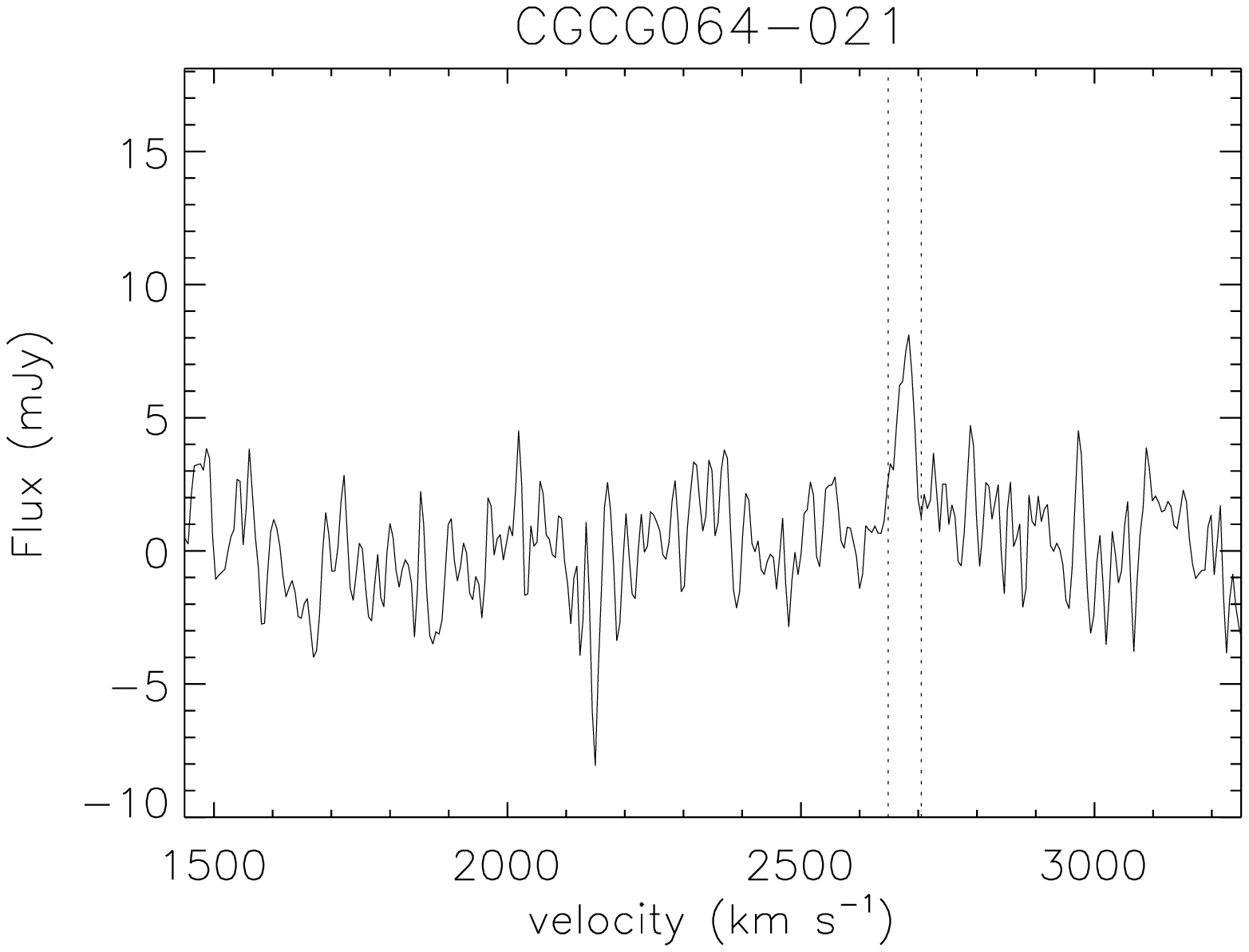} 
\includegraphics[width=4.5cm]{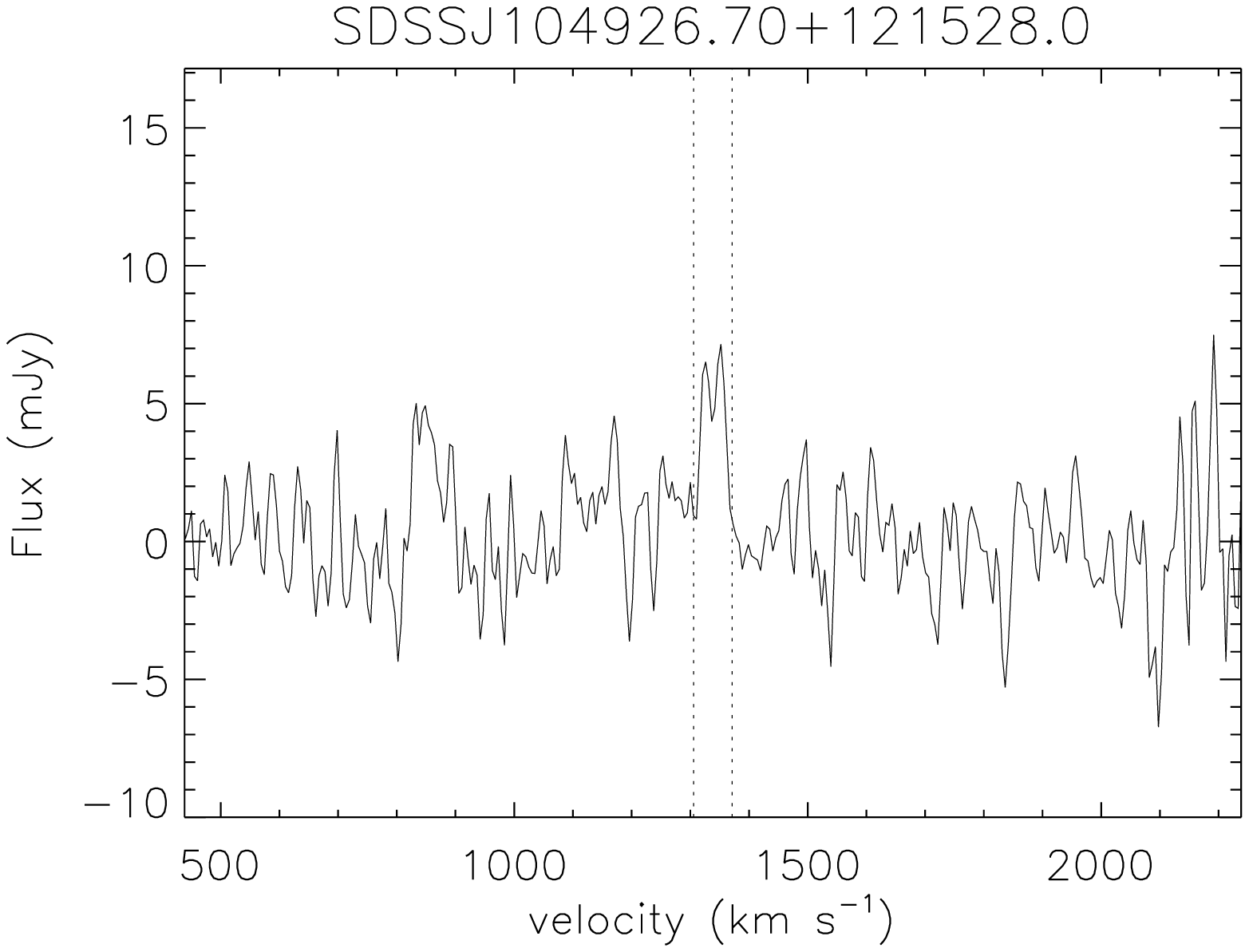}
\includegraphics[width=4.5cm]{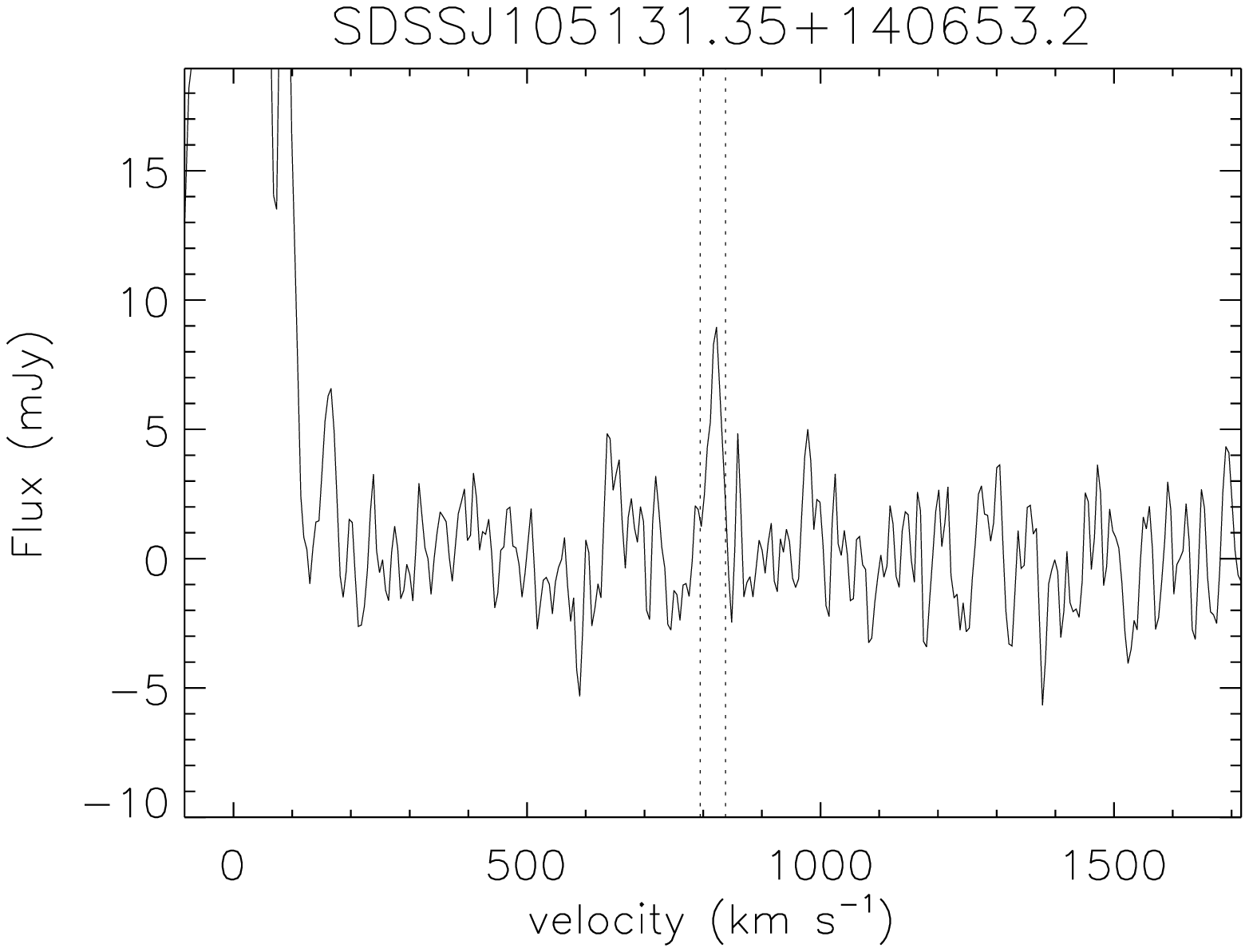}
\includegraphics[width=4.5cm]{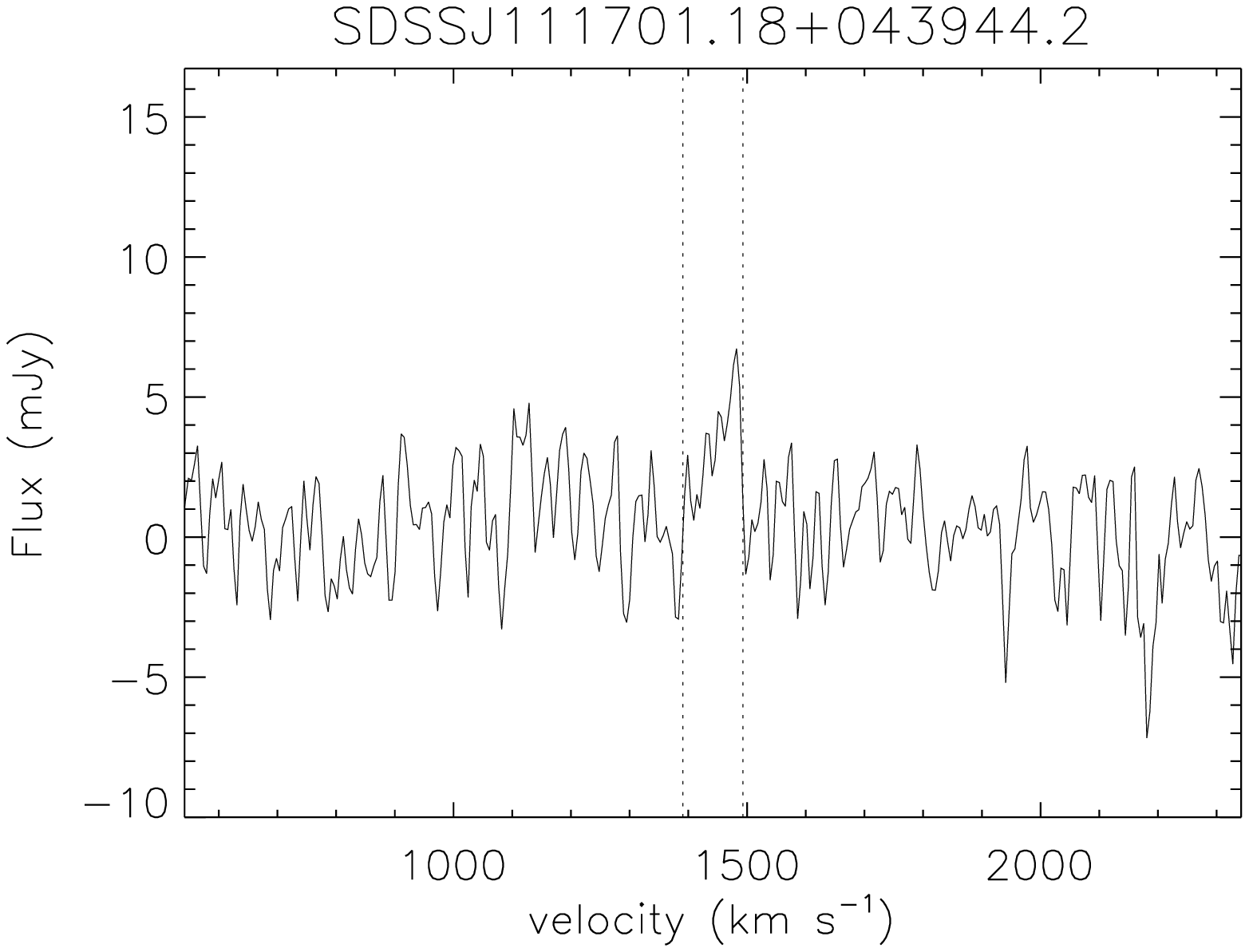} 
\includegraphics[width=4.5cm]{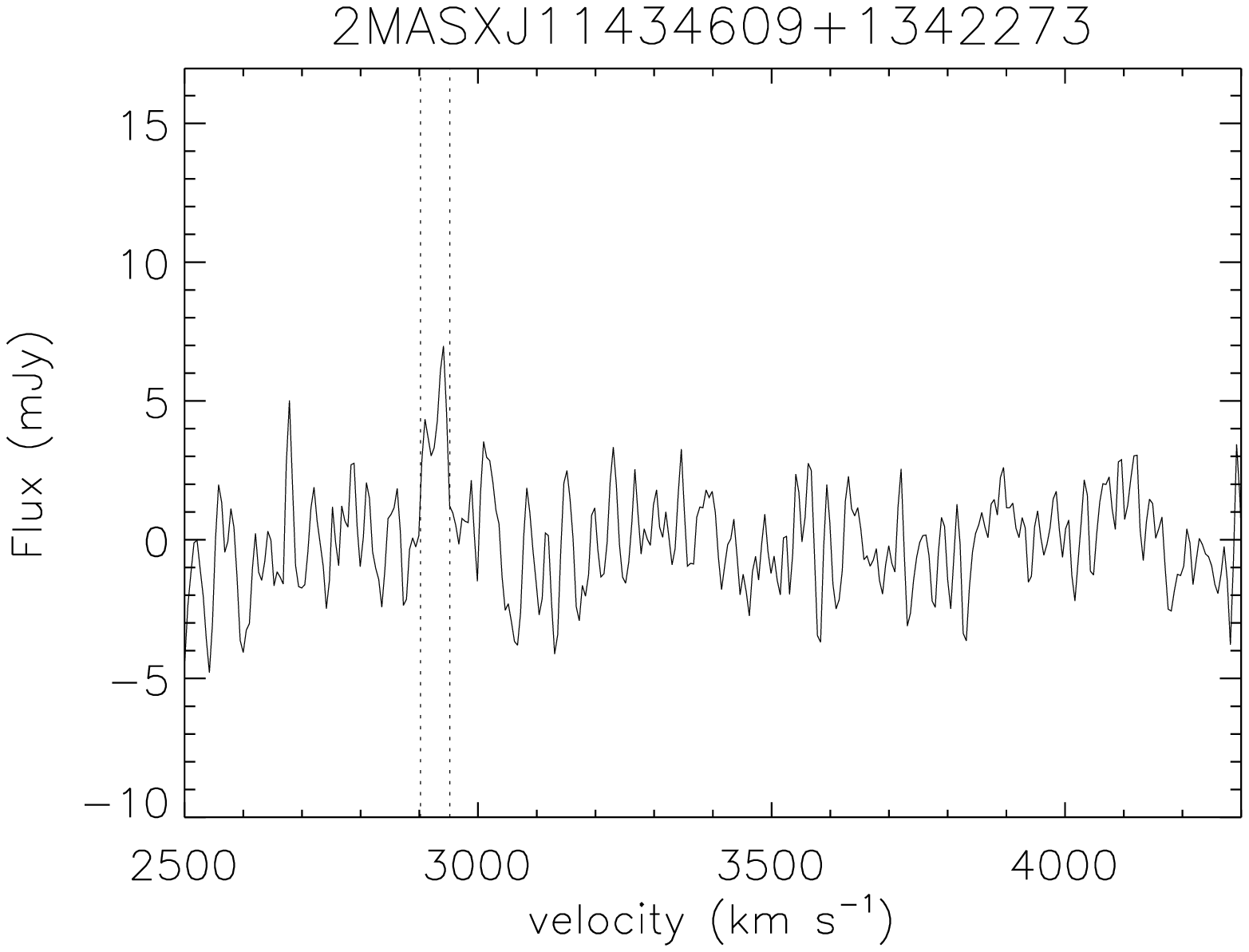} 
\caption{21-cm spectra of the ETGs with \hi emission. The last five spectra correspond to the uncertain detections (code $c$ in Table 1). }
\label{hispec}
\end{center}
\end{figure*}

\begin{figure*}
\begin{center}
\includegraphics[width=4.5cm]{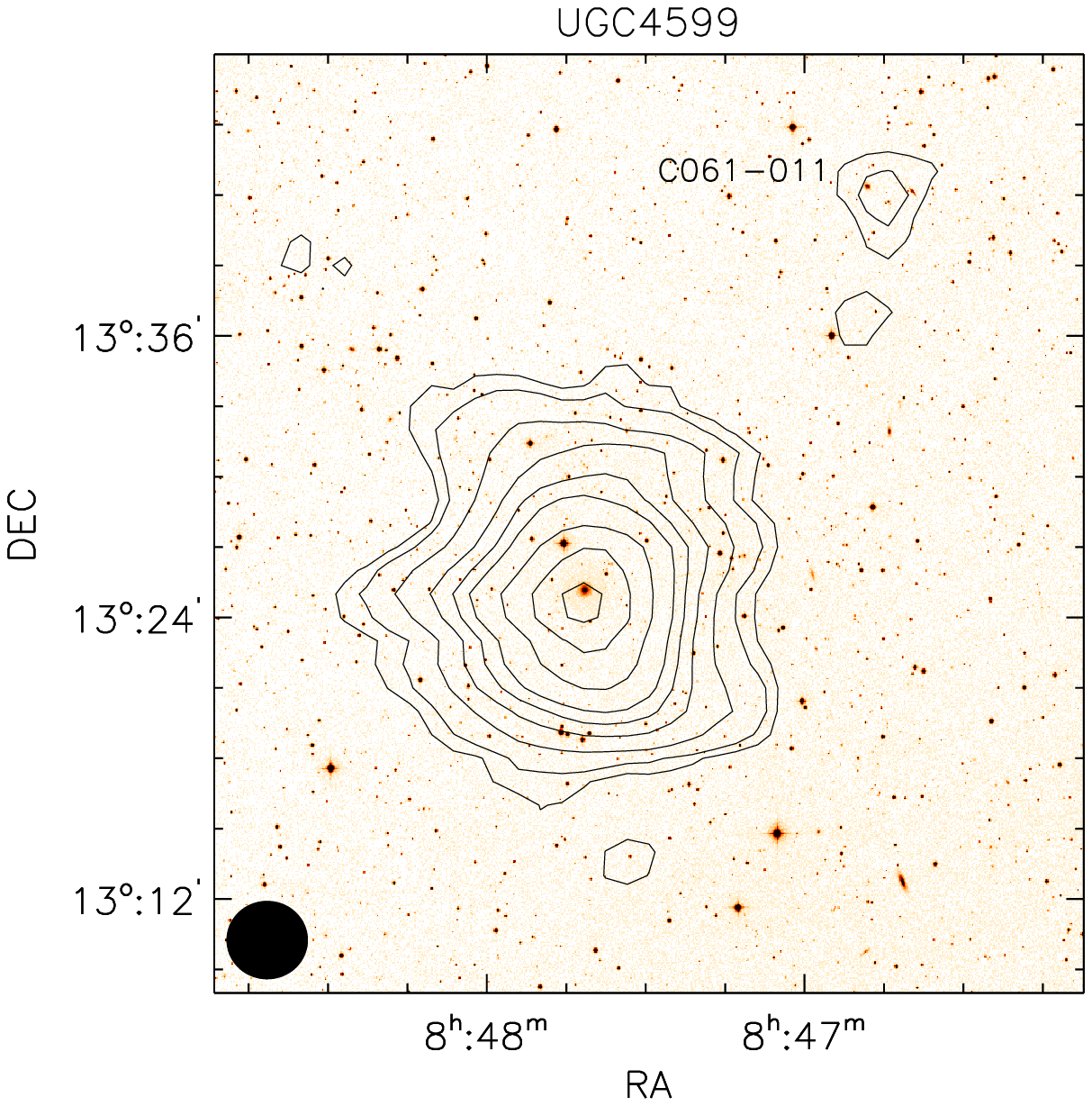} 
\includegraphics[width=4.5cm]{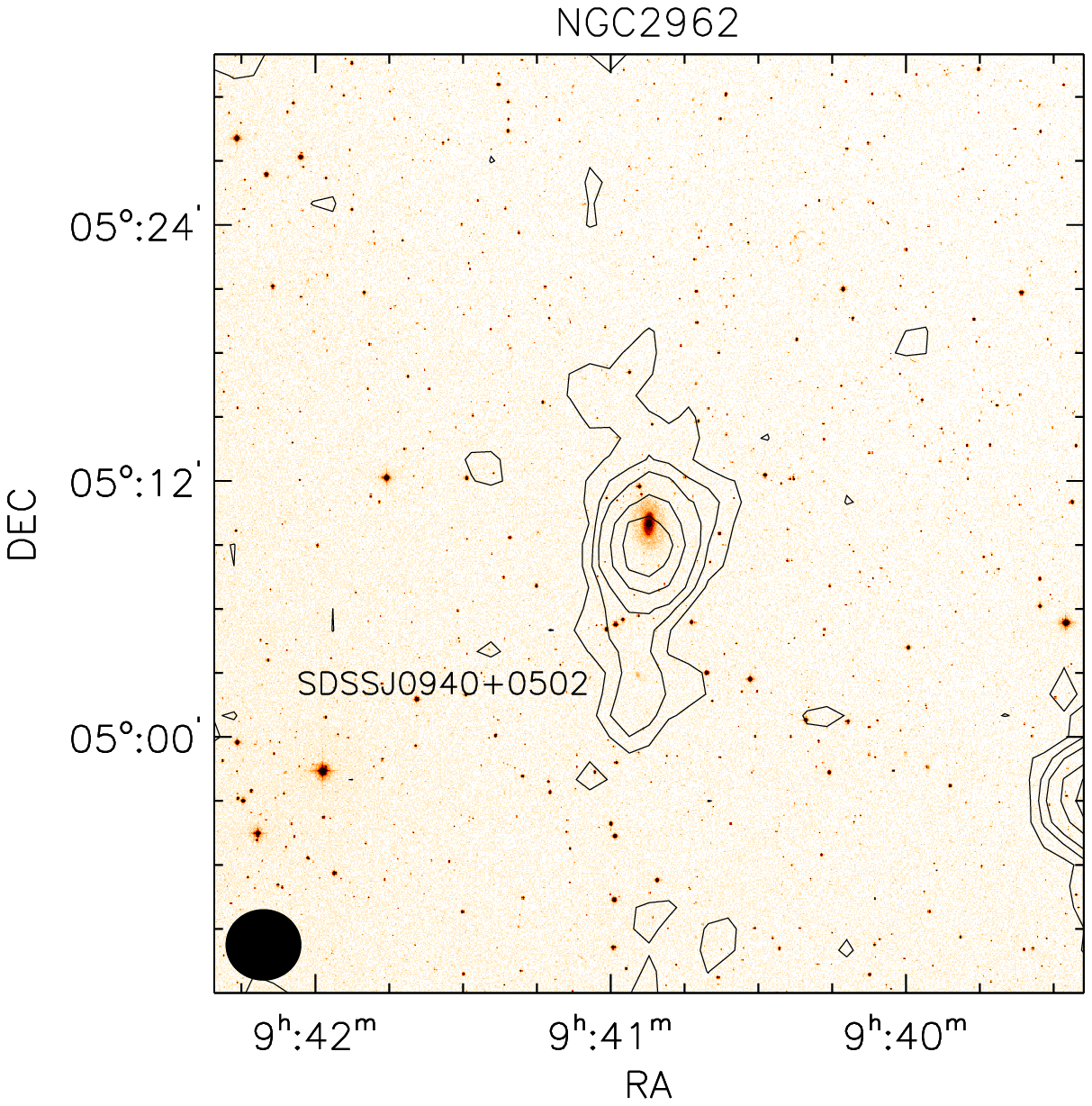} 
\includegraphics[width=4.5cm]{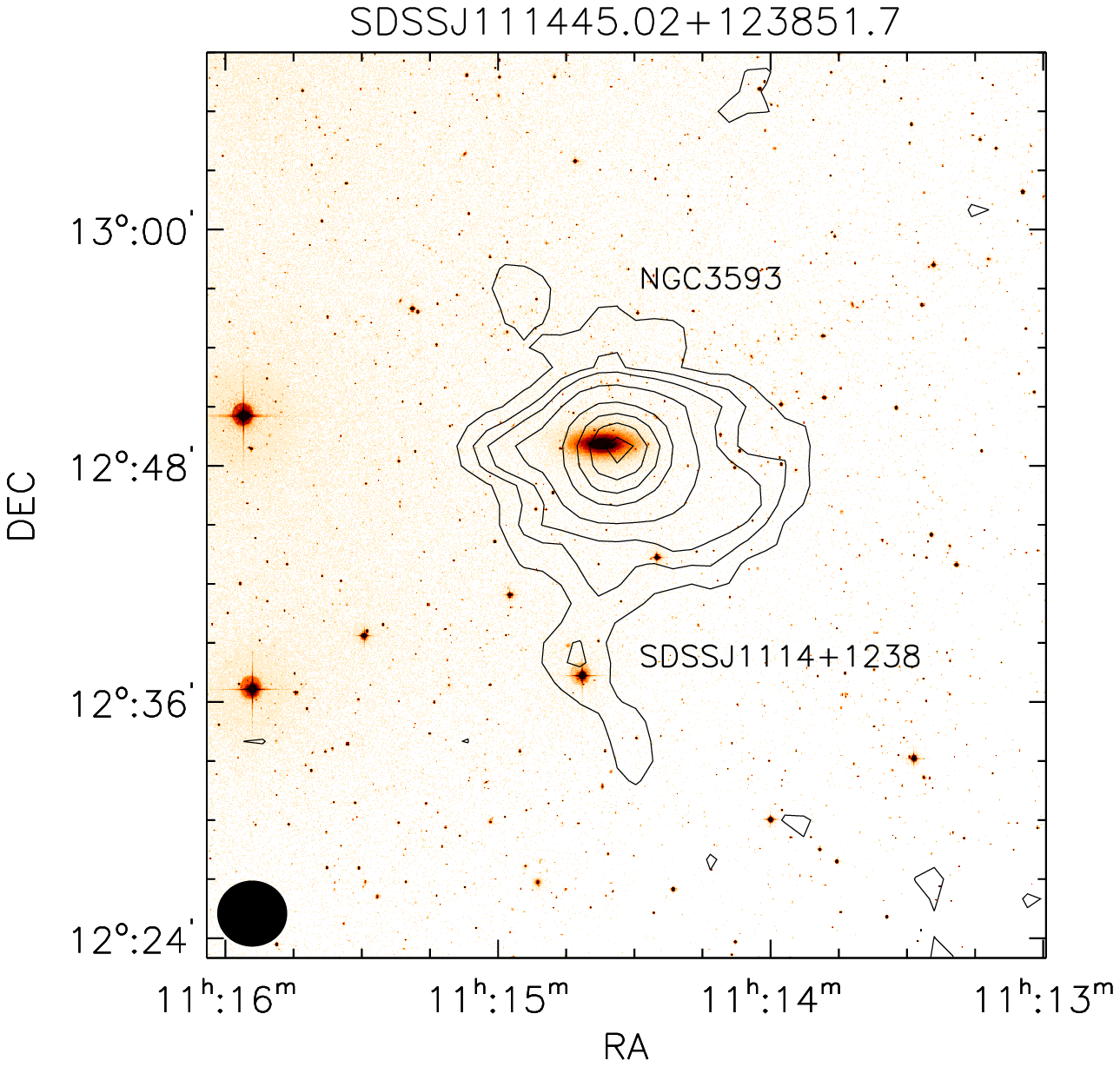} 
\includegraphics[width=4.5cm]{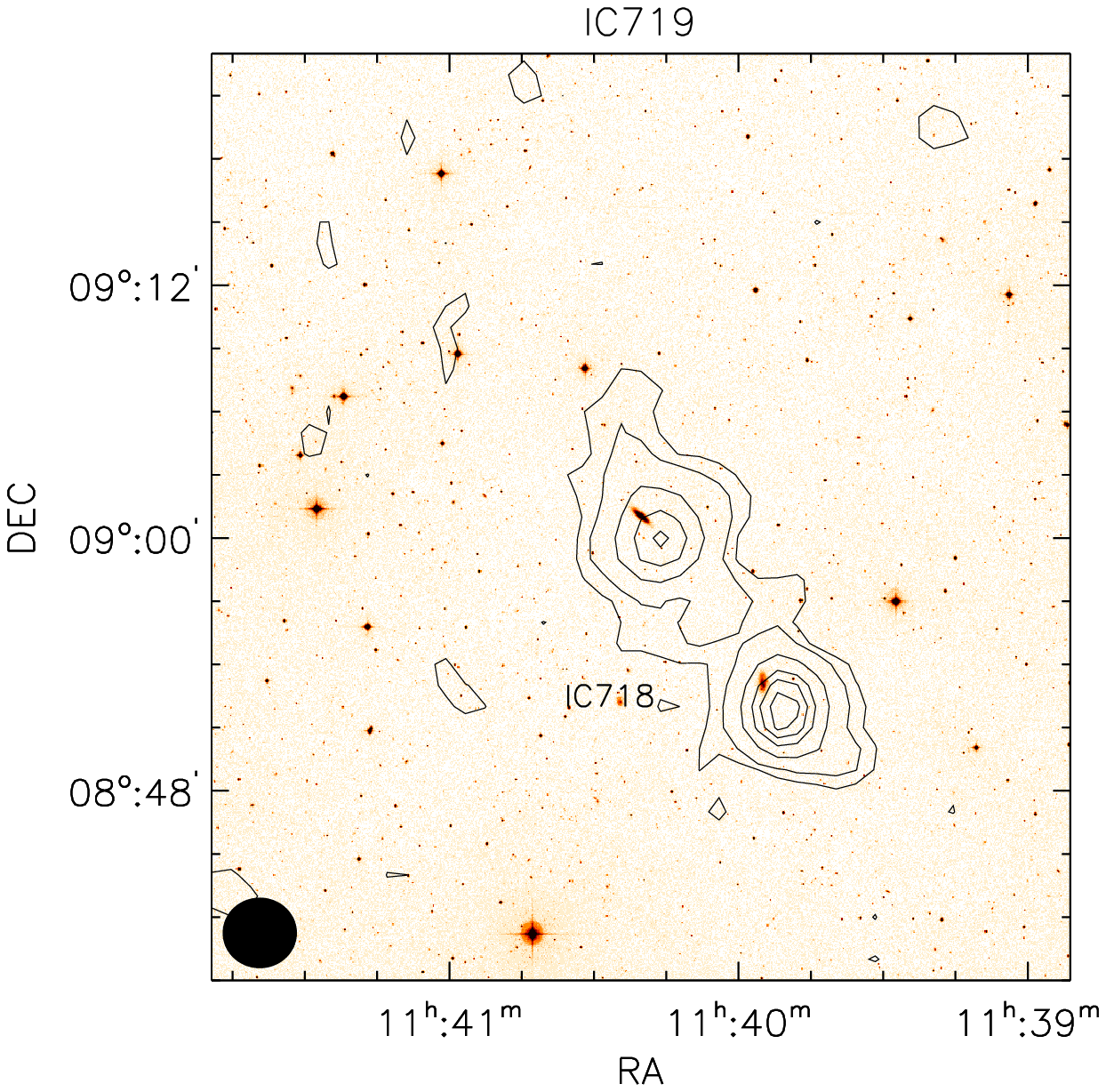} 
\caption{Contour maps of the ETGs with the brightest and most extended \hi emission. The lowest contour level of the maps is at $6 \times 10^{18}$ \cmsq, while the peak column density ranges between 5 $\times 10^{19}$ \cmsq  and $3 \times 10^{20}$ \cmsq. For these galaxies the Arecibo beam ($\sim 4^{\prime}$), which is shown at the bottom left corner of each map, corresponds to a projected linear size which varies between 11 kpc and 40 kpc.}
\label{contours}
\end{center}
\end{figure*}

\section{Results}

\subsection{\hi properties}

For each galaxy we have analysed the ALFALFA grids searching for 21-cm emission.
We found  15 good quality detections out of 62 galaxies ($\sim$25\%; code $a$ and $b$), while five additional candidate sources have lower S/N ($\sim 4$) and they will need deeper observations to be confirmed (code $c$).

15 objects of our sample of 62, have been previously observed at 21-cm.
In particular,
6 out of our 15 detections had previous measurements, and the spectra we
obtain are in good agreement with those in the literature.
Upon inspection of the published spectra we find ours to be of
comparable or better quality with the exception of NGC3489 \citep{1992ApJ...387..484B}, IC719  \citep{
1983AJ.....88.1719G}, and NGC5701
\citep{1996AJ....112..937D}, which are instead of sensitivity higher than ours.
Two others have been previously
detected but with a marked disagreement with our data;
this is the case of UGC4599 \citep{2007A&A...465...71T, 2000ApJS..130..177R},
and of IC676 \citep{2004A&A...416..515D}.
In both cases our data are definitely of superior quality.
As for the undetected ones, 4 objects have in the literature
\hi limits worse than those here provided (NGC3379, 3412, 4880, 5770),
while for NGC3377 the published limits are lower than ours
\citep{1979ApJ...234..448K}.
We do not confirm the dubious detection of NGC6014 of comparable sensitivity to ours  \citep{1983AJ.....88.1695L}. Finally, \citet{2006ApJ...644..850S} report a 21-cm detection of NGC3384 (M$_{HI} = 1.8 \times 10^6$ \msun) at a level well below our sensitivity.

In total we added 7 new 21-cm detections and, possibly, 5 more sources (code $c$),
doubling (at least) the number of ETGs with neutral hydrogen in this region of the sky.

The spectra of all the \hi detections (including the dubious ones) are shown in Fig. \ref{hispec}.
The atomic gas is regularly rotating in UGC4599 and NGC2962, as indicated by the double horn profiles.  Four galaxies (UGC4599, NGC2962, IC719, and  SDSSJ111445.02+123851.7)  have gas-rich neighbours to which they appear to be connected by a \hi bridge, and their contour maps are displayed in Figure. \ref{contours}. The maps of UGC4599, NGC2962 and IC719 
show that the gas extends far beyond the optical disc in these objects.
Here we discuss the \hi properties of some of the most interesting 21-cm detections.

{\bf UGC4599.} It is the galaxy with the highest \hi mass of the sample  (M$_{HI} = 7.6 \times 10^9$ \msun). The gaseous distribution is very extended ($\sim$ 100 kpc) and in rotation as indicated by the double horn 21-cm profile. It belongs to the compact group n. 79 in the Updated Zwicky Catalog  \citep[UZC;][]{2002A&A...391...35F}. From the ALFALFA cube one can see a possible companion also detected at 21-cm (CGCG061-011) with V$_{hel}$ = 2087 \kms (Fig. \ref{contours}). In a few channels there is evidence for \hi emission in the region between the two galaxies which connects to the disc of UGC4599, although this feature is not visible in Fig. \ref{contours}.

{\bf NGC2962.} The \hi emission  appears to be connected to that of the galaxy SDSSJ094056.3+050240.5, being located at a projected distance of about 8 arcminutes ($\sim 80$ kpc; see Fig. \ref{contours}).  In the Lyon Group of Galaxies catalog \citep[][]{1993A&AS..100...47G} it belongs to the group n. 178 whose brightest member is NGC2966. From the ALFALFA data, within a box of 1.5 degrees in RA and DEC and a radial velocity range of $\pm 250$ \kms, there are 8 systems which are detected at 21-cm. The  rotating \hi structure shows the largest velocity width of the sample ($\Delta$V$_{50} = $ 415 \kms).


{\bf NGC3489.} This lenticular galaxy belongs to the Leo Group, but it is in between the two subgroups defined by M66 and M96. It is among the faintest detection with one of the lowest \hi masses (2 $\times 10^7$ \msun). The \hi structure is not resolved by the Arecibo beam thus it is not shown in Fig. \ref{contours}.

{\bf SDSSJ111445.02+123851.7} A dwarf  galaxy in the M66 subgroup of Leo.
The 21-cm emission (Fig. \ref{contours}) extends  to the north towards the brighter  companion NGC3593 (V$_{hel} =$  628 \kms;  at a projected distance of 10 arcminutes/30 kpc), a S0a galaxy  with two counterrotating  stellar discs of different scale length and surface brightness \citep{1996ApJ...458L..67B}. The smaller and less massive one corotates with the \hi gaseous disc, contrary to the more massive stellar disc. The accretion of a gas-rich satellite is addressed as a possible mechanism to explain the double structure of NGC3593 \citep{1999ASPC..163..292P}.


{\bf IC719.} It is a S0 galaxy fairly isolated. It shows only one companion (IC718; V$_{hel} =$ 1860 \kms) to which it seems to be connected by a \hi bridge (see Fig. \ref{contours}).



\subsection{\hi content vs. B band luminosity}

\begin{figure}
\begin{center}
\includegraphics[width=8.5cm]{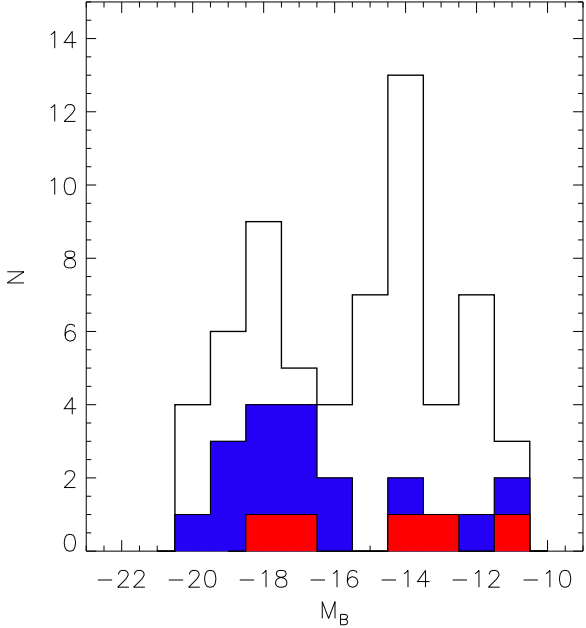} 
\caption{The absolute magnitude distribution of the ETGs detected at 21-cm (filled histogram), compared to the total number of galaxies in our sample. The 5 uncertain \hi detections are shown in red (light grey).}
\label{Abs_mag}
\end{center}
\end{figure}

The SDSS database provides $u,g,r,z,i$ photometry for all the galaxies of our sample, including those for which spectroscopic information is not available. As mentioned in Section 2, B apparent  magnitudes have been taken from the RC3 for the galaxies included in this catalog, otherwise they have been computed from the dereddened SDSS photometry using the relation from Smith et al. (2002).
Figure \ref{Abs_mag} displays the absolute magnitude distribution of all the galaxies, compared to those detected at 21-cm (filled histogram). The five code $c$  detections are also shown.
Given that our sample is complete down to $r < 17.7$ we can evaluate what is the completeness magnitude limit in M$_B$. The $B-r$ colour distibution of these galaxies ranges between 0.4 and 1.4 and the maximum distance modulus is $m-M = 33.18$; this implies that the sample we have selected is at least complete down to M$_B= -15$.

A large fraction of the observed luminous (M$_B < -17$) early-type systems (10 out of 23, 44\%) contain neutral hydrogen. At fainter magnitudes (M$_B > -17$) we find 5 out 39 gas-rich dwarfs (13\%, not taking into account the uncertain detections), although the incompleteness of the sample becomes important for M$_B > -15$.
The detection rate per magnitude bin peaks at M$_B = -17$ (60\%), and then it drops at magnitudes fainter than M$_B = -16$.

Other studies in the field find similar high \hi detection rates in ETGs, however the difference with a dense environment like Virgo is striking. In Paper I we found  that only 2 out of 55  galaxies  brighter than M$_B = -17$ have neutral hydrogen (about 4\%). The fraction of luminous ETG with \hi in LDEs is 10 times higher than in Virgo. However the three brightest  ellipticals  of the LDE sample (NGC3377, NGC3379 and NGC3384)
do not show \hi emission down to the sensitivity limit of the ALFALFA data set (see Table 1).
On the fainter side of the distribution, the \hi detection rate for dwarf ETGs in LDEs is almost ten times higher than in Virgo where we found only 7 gas-rich early-type dwarfs out of 407 (1.7\%, see Paper I).

\subsection{The \ml distribution of the sample}

\begin{figure}
\begin{center}
\includegraphics[width=8.5cm]{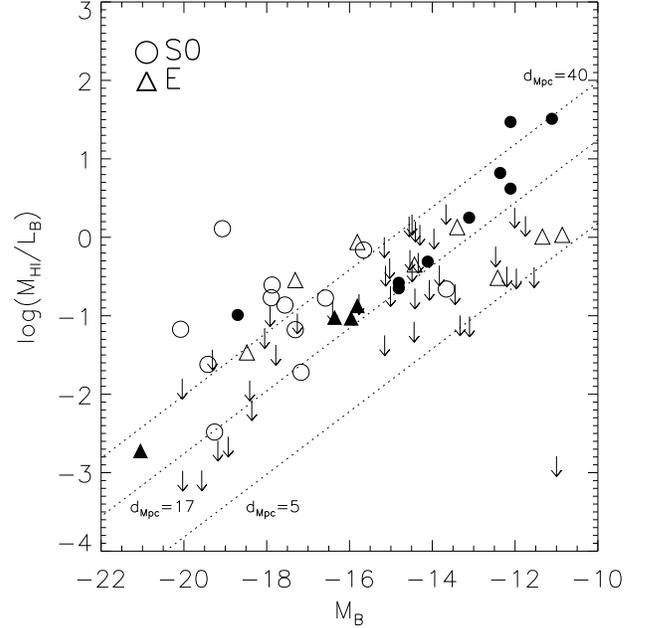} 
\caption{The gas mass-to-light ratio versus the absolute magnitude of the LDE sample  (open symbols) compared to the Virgo one analysed in Paper I (filled symbols). Arrows indicate the upper limits for the galaxies non-detected at 21-cm. The dotted lines show the 5$\sigma$ ALFALFA detection limits at three different distances (d=5, 17, 40 Mpc), assuming a rms of 2 mJy, and a velocity width of 80 \kms.   }
\label{ML}
\end{center}
\end{figure}

In Figure \ref{ML} we compare the gas-mass-to-light ratios of the LDE and Virgo cluster samples displayed as a function of the absolute B magnitude.
The \hi mass-to-light-ratios (\ml) for the LDE sample range between 3 $\times 10^{-3}$ (NGC3489) and 1.3 in solar units (UGC4599 which shows an unusually high \ml  for a S0 galaxy).
Non detections are indicated with their upper limits on \ml.
As one can notice from Fig. \ref{ML} there is a spread in the \hi content of the luminous ETGs (M$_B < -17$), and the \ml ratios of the sample  do not seem to be correlated with the luminosity, at least at brighter magnitudes, 
as it is generally found in other morphological types \citep{2001A&A...370..765V,2008A&A...482...43G}. This could give an indication that gas-rich luminous objects have recently accreted gas  either via merging or from the IGM as it has been suggested in previous works \citep{1985AJ.....90..454K}. The figure clearly shows the difference with respect to Virgo, where only two objects with M$_B < -17$ have been detected at 21-cm, and the only \hi detected luminous elliptical (M86) has a very low \ml ratio.

From the \ml distribution of the sample we can check whether we are finding a a "peculiar" set of galaxies with a  higher gas content than what is usually expected for these morphological types.
The median \ml for ellipticals and S0s in the Uppsala General Catalog (UGC) is 0.04 \citep{1994ARA&A..32..115R}.
For the luminous galaxies in our LDE sample we find \ml = 0.07,  which is roughly comparable to the
UGC.
On the other hand, the dwarf ellipticals/spheroidals detected with ALFALFA have a median gas-mass-to-stellar light ratio, \ml = 0.9, which is more similar to those of dwarf irregulars (\ml $\lesssim 1$).  In the Local Group, for example, the dE galaxies NGC205 and NGC185 have  both \ml = 0.001 M$_{\odot}/$L$_{\odot}$  \citep{1998ARA&A..36..435M}, while
early-type dwarfs in the Sculptor group with \hi masses around 10$^5$ \msun, exhibit \ml ratios between 0.08 and 0.2 \citep{2005AJ....130.2058B}. The lack of low luminosity galaxies with low relative gas content is probably a selection effect due to the smaller volume that the ALFALFA sensitivity allows for these systems at a given \ml ratio, as shown in Fig. \ref{ML}.

\begin{table*}
\caption{$u-r$ colour and  emission-line fluxes of the galaxies with ionised gas. The horizontal line divides \hi detections from non-detections. For the objects not included in the SDSS spectroscopic sample, the fluxes have been taken from the values reported in the literature. The  references are indicated in the last column and they correspond to: 1.  SDSS \citep[]{2002AJ....124.1810S}; 2. \citet{2005ApJ...634..210G}; 3. \citet{1997ApJS..112..315H}; 4. \citet{2006ApJS..164...81M}.}
\begin{minipage}{2\textwidth}
\begin{tabular}{lccrcccccc}
 \hline \hline
           ID  & & $u-r$ &&       H$\beta$ $\: $ &    O[III]$_{5007}$   &  H$\alpha$ $\: $ & N[II]$_{6583}$     & & Ref\\
             & &     &&     $10^{-17}$erg s$^{-1}$cm$^{-2}$   & $10^{-17}$erg s$^{-1}$cm$^{-2}$   &$10^{-17}$erg s$^{-1}$cm$^{-2}$  &$10^{-17}$erg s$^{-1}$cm$^{-2}$  & &  \\
\hline \hline
                  MRK0706  & & 1.62 && 2671 $\pm$ 37 &  5035 $\pm$ 62 &  10850 $\pm$ 125 &   1981 $\pm$ 30    & & 1\\
  SDSSJ093608.59+061525.4  & &  1.41 && 74 $\pm$  4 &    151 $\pm$  6 &    347  $\pm$  7 &   57  $\pm$ 4  & & 1 \\
                  NGC2962  & & 2.85 && --       &   --           &    20       & 130   & & 2\\
  SDSSJ104926.70+121528.0  & & 1.02 && 260 $\pm$  8 &   892 $\pm$  14 &    857 $\pm$  14 &   40 $\pm$  4 & & 1\\
                  NGC3489  & & 3.73 && 1488   &    4367    &     3235           &   4302         & & 3\\
                    IC676  & & 2.00 && 589 $\pm$  17  &    175 $\pm$ 16 &   4071 $\pm$ 57 &  1921 $\pm$ 34 & & 1\\
  SDSSJ111445.02+123851.7  & & 1.27 && 1.5 $\pm$ 1.3   &    10 $\pm$ 3 &     19  $\pm$ 2  &    3.0  $\pm$ 2  & & 1\\
                    IC692  & & 1.23 && 165 $\pm$  21  &   563 $\pm$  23 &    753 $\pm$  9  &  120 $\pm$  11  & & 1\\
   2MASXJ11434609+1342273  & & 1.89 && 832 $\pm$  14  &   641 $\pm$  11 &   4187 $\pm$ 53 &  1286 $\pm$  17 & & 1\\
                  NGC3773  & & 1.13 && (271 $\pm$ 11)$\cdot 10^2$    &   (355 $\pm$ 15)$\cdot 10^2$   &     (879 $\pm$ 35)$\cdot 10^2$        &  (157 $\pm$ 7)$\cdot 10^2$     & & 4\\
                  UGC6655  & & 1.28 && 1635 $\pm$ 31  &  5352 $\pm$ 96 &   5563 $\pm$ 67 &   515 $\pm$  12 & & 1\\
   2MASXJ11460404+1134529  & & 1.22 && 1135 $\pm$  20  &  3294 $\pm$ 43 &   4294 $\pm$ 47 &    424$\pm$  12 & & 1\\
                  NGC5338  & & 1.91 && 407 $\pm$ 15  &   531 $\pm$  18 &    2742 $\pm$  19  &  893 $\pm$  21 & & 1\\
  SDSSJ144329.18+043153.4  & & 1.68 &&  9 $\pm$  2  &     16 $\pm$  2 &     41 $\pm$  2 &     5 $\pm$  1 & & 1\\
\hline
                  UGC4590  & & 2.23 &&   48 $\pm$ 5  &     16 $\pm$  5 &   507  $\pm$ 11  &   176  $\pm$ 9   & & 1 \\
  SDSSJ095036.25+124832.7  & & 1.64 &&  9 $\pm$  5 &     25 $\pm$  5 &     92 $\pm$  3 &   17 $\pm$  3 & & 1 \\
              CGCG064-055  & & 1.34 && 200 $\pm$  11 &  2714$\pm$  42 &   1810 $\pm$  17 &  1507 $\pm$ 34 & & 1\\
  SDSSJ102339.36+123725.6  & & 1.85 &&  17 $\pm$  3 &    33 $\pm$  4 &     94 $\pm$  3  &   18 $\pm$  3 & & 1\\
    MAPS-NGPO\_559\_1243538 & & 1.82 &&  28 $\pm$  4   &   72 $\pm$  5  &    124 $\pm$  3  &    11 $\pm$  3 & & 1 \\
\hline \hline
\end{tabular}
\end{minipage}
\end{table*}

\subsection{Optical emission lines and colours}

To check whether we can find traces of a recent star formation activity and/or differences in the stellar population
in our sample of ETGs, we have looked for optical emission lines in the SDSS spectroscopic database and we have
analysed the optical colours provided by the SDSS photometric database.

Different ionising mechanisms can produce emission line spectra in galaxies, however  the use of diagnostic diagrams \citep{1981PASP...93....5B, 1987ApJS...63..295V}, for given values of the emission line ratios [OIII]$_{5007}$/H$\beta$ and [NII]$_{6583}$/H$\alpha$, allows to distinguish the spectra of normal star forming (SF) systems from those  with an active galactic nuclei (AGN), such as Seyfert galaxies and low-ionization nuclear emission regions (LINERs) \citep[][]{2001ApJ...556..121K,2003MNRAS.346.1055K,2006MNRAS.371..972S}.

\begin{figure}
\begin{center}
\includegraphics[width=9.5cm]{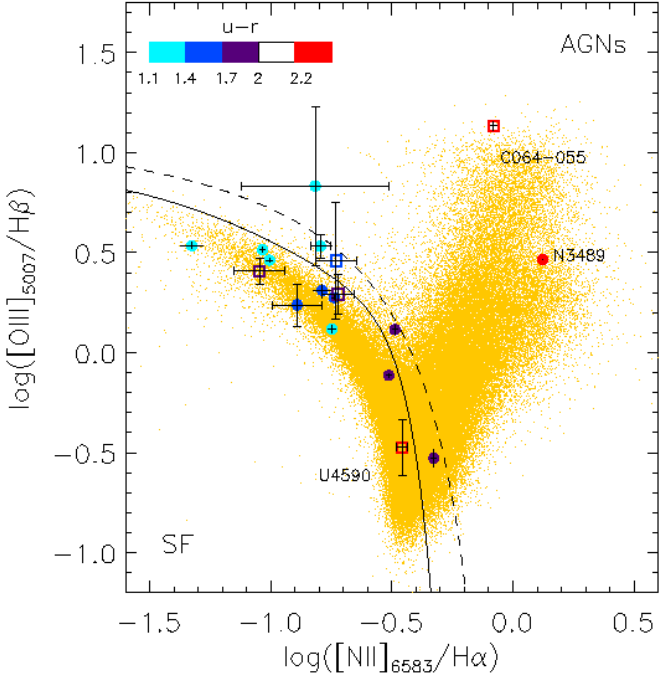} 
\caption{The [NII]$_{6583}$/H$\alpha$ vs. [OIII]$_{5007}$/H$\beta$ line ratios for the galaxies in our sample with emission line spectra. Filled circles correspond to galaxies with \hi, and open squares to 21-cm non detections.  The bar at the top left corner of the plot shows the range of the $u-r$ colour of the galaxies. The curved lines represent the demarcation between SF systems and AGNs according to \citet{2006MNRAS.371..972S} (solid line) and \citet{2003MNRAS.346.1055K} (dashed line). The small gray dots give the distribution of the SDSS DR5 sample of \citet{2008MNRAS.391L..29S}, clearly showing that galaxies occupy two well defined regions where AGNs (Seyferts and LINERs on the righthand side) can be distinguished from SF galaxies (the so-called "left wing of the seagull").
}
\label{diagnostic}
\end{center}
\end{figure}

We have extracted from the SDSS database the fluxes of the four lines H$\beta$, [OIII]$_{5007}$, H$\alpha$ and [NII]$_{6583}$ (Table 2) and their ratios are shown in Figure \ref{diagnostic}. For the 12 galaxies without SDSS spectra we have used the values reported in the literature when emission lines had been observed, as in NGC2962 \citep{2005ApJ...634..210G}, NGC3489 \citep{1997ApJS..112..315H} and NGC3773 \citep{2006ApJS..164...81M}.
The solid and dashed lines in the figure define the region of the diagram below which SF galaxies can be found, according to Stasinska et al. (2006) and Kauffmann et al. (2003) respectively.
 10 out of the 15 certain 21-cm detections fall in the SF region of the diagram (67\%). As regards the remaining five galaxies, line ratios of NGC3489 indicate that this is a Seyfert galaxy \citep{1997ApJS..112..315H}, while  the other four have no clearly detected [OIII] emission lines. Particularly, the spectrum of NGC2962 does not show neither [OIII]$_{5007}$ nor H$\beta$ emission lines \citep{2005ApJ...634..210G}, as well as IC719, thus we do not display these objects in Fig. \ref{diagnostic}, even though the [NII]$_{6583}$/H$\alpha$ ratio would place them in the AGN region of the diagram.
The SDSS spectrum of IC2684 shows low S/N H$\alpha$, and [NII] emission lines, thus it is difficult to draw any conclusions on the position of this galaxy in the diagnostic diagram.
Finally, the spectrum of  UGC4599, which is the galaxy with the highest \hi mass, have no emission lines, but for the latter galaxy one has to take into account that
the SDSS spectrum is sampling only its inner 3$^{\prime\prime}$ region, while the multi-colours image shows a blue ring where SF is probably occurring at about 30$^{\prime\prime}$ from the centre.
Two of the five uncertain \hi sources (SDSSJ104926.70+121528.0 and 2MASXJ11434609+1342273) have emission line ratios typical of H{\sc ii} regions.

It is interesting to note the presence of four ETGs non detected at 21-cm in the SF area of the diagram (UGC4590, SDSSJ102339.36+123725.6, MAPS-NGPO\_559\_1243538 and SDSSJ095036.25+124832.7). The spectrum of CGCG064-055 instead appears  to be typical of a Seyfert galaxy.

$g-r$ versus $u-g$ colour-colour diagrams are used to separate between early and late-type morphology in the SDSS (Strateva et al. 2001, Lamareille et al. 2006) and the boundary between the two classes of galaxies is set at $u-r = 2.2$ (Strateva et al. 2001). In Fig. \ref{diagnostic} we display the
$u-r$ colours of the galaxies with emission lines. One can notice that all the systems detected at 21-cm falling in the SF region have blue colours  ($u-r < 2$), more similar to those of later morphological types (Lamareille et al. 2006); the bluer the galaxy the greater the [OIII]$_{5007}/$H$_{\beta}$ ratio.
As regards the ETGs non detected at 21-cm in the SF region, they also  have colours bluer than $u-r = 2$, except for UGC4590, whose spectrum shows weak [OIII] emission lines.
The reamaining galaxies outside the SF region both with or without \hi such as NGC3489, CGCG064-055 and NGC2962 (not shown in the plot) have all colours $u-r > 2.2$ typical of early-type morphologies.

\section{Discussion}

\subsection{The role of the environment}

The aim of our study  was to assess the extent to which the neutral hydrogen content of ETGs can be affected by the environment. 
In this paper we have selected a uniform set of galaxies which are members of
groups (both small and more populated such as Leo), or evolve in relative isolation, and we have compared it to a rich cluster like Virgo (Paper I).

The reduced \hi content of early-type systems in Virgo is remarkable when compared to the LDE sample.
The percentage of galaxies detected at 21-cm is 10 times higher in regions of low intergalactic density both for luminous and dwarf ETGs, confirming that the impact of the environment on the gaseous content of galaxies is fundamental.

Several factors may concur to the higher detection rate of \hi in LDEs. 
Contrary to clusters, groups have low velocity dispersions and crossing times smaller than the Hubble time, so that interactions and mergers are more likely to occur 
\citep{2007ggnu.conf..203M}.
The lack of an intragroup hot medium and the low relative velocities imply that the role of ram pressure stripping \citep{1972ApJ...176....1G} is reduced in LDEs.
On the other hand, luminous ETGs in Virgo are primarily confined to the central regions of the cluster and are strongly concentrated around M87 \citep{1987AJ.....94..251B},
where mechanisms  such as ram-pressure, or turbulent/viscous stripping \citep{1982MNRAS.198.1007N} and tidal interactions \citep{1996Natur.379..613M} are more efficient at removing  the gas from a galaxy.

Another explanation for the paucity of gas-rich luminous ETGs in Virgo compared to the LDE case, may be the relatively short evaporation time for a cold gas cloud embedded in a hot intracluster medium. Following \citet{1977ApJ...215..213M}
we estimate that the evaporation time due to heat conduction for a cloud of $\sim$ 500 pc radius and $n \sim 1$ cm $^{-3}$ density, embedded in a hot plasma
with properties typical of Virgo \citep{2004ApJ...609..638G} is around few times $10^8$ yrs (for
conductivity at the Spitzer's value). 
Within this scheme the paucity of \hi detections of bright cluster ETGs compared to lower density environments derives from the dearth of \hi rich dwarfs to be cannibalized. 
Within 6$^{\circ}$ from M87, we estimate from the VCC \citep{1985AJ.....90.1681B} an average density of gas-rich dwarfs $n_{HI}^d \simeq 10$ gal Mpc$^{-3}$ with a velocity dispersion $\sigma \simeq 700$ \kms. If we fix a cross section diameter of 100 kpc for the capture by a giant member, this implies an impact rate of $\sim 5.3 \times 10^{-2}$ Gyr$^{-1}$. That is, in each bright ETG, the time between collisions/capture of a \hi rich dwarf amounts to $\sim$ 19 Gyr. Given, as said before, a survival time of the neutral gas of $\sim$ 6 $\times$ 10$^8$ Myr, 
this gives an expected detection rate around 3\%, 
in agreement  with the observed one.

Finally an important role in evaporating \hi clouds could be also played by the possible presence of a hot ISM in luminous ETGs. At this stage, unfortunately, we can derive little information
about the relation between the cool and hot ISM for the LDE sample, since the
overlap with  X-ray datasets is very small.
Few objects have been observed, but
no evidence of a hot ISM has been reported in any of them to relatively
low levels \citep{2001MNRAS.328..461O, 2006ApJ...653..207D}, with the exception
of NGC3379, where a small amount of hot gas is detected in an outflow
phase \citep{2008ApJ...688.1000T}.
A more systematic analysis of the effects of a hot plasma on the \hi content of ETGs in different environments will be presented in a subsequent work (Grossi et al., in prep.).

\subsection{Star formation in LDE ETGs}

Warm ionised gas is considered a reliable tracer of star formation activity in spiral galaxies, but in
E and S0 galaxies other ionising sources such as AGNs or old stars (post asymptotic giant branch stars or low mass core helium-burning stars) can produce emission line spectra \citep{1990ASSL..160..301D,1991AJ....101.1647T,1994A&A...292...13B}. The presence of ionised gas is quite common in ETGs \citep{2006MNRAS.366.1151S}, particularly \citet{2006NewAR..49..515F} find that emission lines are slightly more frequent in S0s (83\%) than in ellipticals (66\%). In general, line ratios in ETGs, especially in the more luminous ones, are typical of LINERs \citep{1986AJ.....91.1062P,2003ASPC..290..369F,2008A&A...483...57S}.

We find that 40\% of the ETGs in our sample have H$\alpha$ equivalent widths (EWs) greater than 1 $\AA$, indicating the presence of ionised gas (see Table 1). This fraction is lower than what is generally found in other samples of ETGs.

Searching for optical emission lines other than H$\alpha$ we find that 30\% of our ETGs have spectra with H$\beta$, [OIII] and [NII] in emission. The majority of them lie in the SF side of the diagnostic diagram displayed in Fig. \ref{diagnostic}, the so-called 'left wing of the seagull' \citep{2006MNRAS.371..972S}. The upper side of the left wing corresponds to low metallicity SF galaxies, while more massive and metal-rich galaxies populate the bottom of the SF region. Objects move to the right of the left wing as the metallicity increases. Combining the presence of the optical lines with the colours from the SDSS photometry, 12 out of 20  ETGs with neutral hydrogen (60\% if we include code $c$ detections) appear to show traces of ongoing SF activity.
The colours of the galaxies in the SF region are predominantly blue and particularly they are bluer than $u-r=2.2$, values that are more similar to those of later type morphologies. Thus a significant fraction of the ETGs with gas seems to be forming stars.

If we exclude IC676, all the gas-rich ETGs in the SF region are either low luminosity or dwarf ETGs with M$_B > -18$. Evidence of ongoing star formation activity is not unusual in the central regions of low luminosity ETGs \citep{2000AJ....119.1180S}. Semi-analytic models expect  that
low luminosity ellipticals and field ellipticals may have experienced more extended star formation histories or could have been formed in later mergers \citep{2001ApJ...561..517K}.

Even though star formation could be the dominant source of ionisation in the ETGs of our sample with optical emission line spectra, one has to consider that the diagnostic diagram does not allow to unambiguously distinguish whether the gas is ionised by massive young stars or hot old stars \citep{2008MNRAS.391L..29S,2008MNRAS.384.1045K}, and a further analysis of the UV colours is needed to disentangle this issue.

Finally 35\% of the firm \hi detection galaxies (or 40\% if one includes the code $c$ objects) do not present emission lines in their optical spectra, or do not have line ratios typical of H{\sc ii} regions. A possible explanation could be that star formation has now faded, but it occurred in a recent past ($< 1$ Gyr). Indeed some of these gas-rich galaxies have relatively blue colours for their moprhological type (such as IC2684 with $u-r = 2$).
Alternatively, the lack of SF activity in ETGs with gas could be due to the too low column densities  of the observed gaseous structures in their sample (at most a few times $10^{20}$ \cmsq), which would not allow to trigger star formation on a large scale. Thus it is possible that, even if ETGs host a gaseous component, only in some of them the conditions for star formation are attained and the gas converted into stars. If this does not occur, the gaseous discs and rings could survive for several Gyr \citep{2006MNRAS.371..157M,2007A&A...465..787O}.

\section{Conclusions}

To better understand the environmental effects on the evolutionary path of ETGs,
after having studied the \hi content of a sample in a dense cluster like Virgo (Paper I) here we have used the ALFALFA data set to make a similar analysis  on a LDE sample to identify any differences in the ISM and stellar properties of such galaxies. Using the SDSS database we have built an apparent magnitude limited sample of 62 ETGs with radial velocity below 3000 \kms which is complete down to $r < 17.77$ mag.

Our main results are:

\begin{itemize}

\item The \hi detection rate of luminous ETGs in LDEs is 44\%, 10 times higher with respect to what we found in the Virgo cluster. 
    The three brightest ellipticals  (the Leo group members NGC 3377, NGC 3379 and NGC3384) do not appear to have a gaseous component down to the 5$\sigma$ detection limit of $M_{HI} \sim 2 \times 10^7 (\Delta$V/200 \kms)$^{\frac{1}{2}}$ ($d$/11 Mpc) \msun. In agreement with previous studies \citep{1983AJ.....88.1719G} we confirm that the probability of finding S0  galaxies with gas is higher in LDEs compared to a rich cluster like Virgo. The non  detection of ellipticals brighter than M$_B < -20$  at 21-cm prevents to make any comparison  of their \hi properties between different environments.

\item The \hi detection rate lowers to 13\% when one considers only the early-type dwarfs of the sample.  The gas content of the dwarfs (\ml $> 0.2$) appears to be greater than typical values usually found in dE/dS0 morphological types. The fraction of gas-rich dwarf ETGs in LDEs is about 10 times higher than the Virgo cluster.

\item 
    The presence of \hi in LDE ETGs  is possibly the result of recent minor or major merging of gas-rich companions. We find that  $\sim$25\% of the galaxies with \hi emission do also appear to have gas-rich companions. However we cannot exclude that,  given the less 'hostile' environment where such galaxies are evolving in, with gas removal mechanisms being less effective, some of them may have been able to retain their original gas content compared to their cluster counterparts.

\item From the analysis of the emission line ratios of these galaxies  we find that 60\% of the ETGs with gas show traces of ongoing star formation and colours suggesting the presence of young stellar population ($u-r < 2$). The presence of gas seems to be related to a recent star formation activity in a significant fraction of the gas-rich ETGs. In some ETGs, particularly the massive ones, the gas could be ionised by hot old stars.

\item Gas-rich ETGs without traces of recent star formation are also found. They may have not experienced any recent merging/interaction event and they have been passively evolving because the surface densities of their \hi discs are too low for star formation to occur.
    Alternatively we cannot exclude that they may have got their gas from IGM accretion, a process which not necessarily may trigger star forming activity in the galaxy  (Morganti et al. 2006, Oosterloo et  al. 2007).

\item Few ETGs non detected with ALFALFA show emission line spectra of star forming galaxies, and colours indicative of young stellar populations. Deeper 21-cm observations are needed to set stronger constraint on the \hi content of these galaxies.

\end{itemize}

\begin{acknowledgements}
We want to thank Maurizio Salaris, Marcel Clemens, Luc Binette and Grazyna Stasi{\'n}ska for useful discussions, and Natalia Vale Asari who kindly supplied the SDSS DR5 catalog used in Figure \ref{diagnostic}.\\
RG, MPH acknowledge partial support from NSF grants AST 0307661 and AST 0607007, and from the Brinson Foundation.
This work is based on observations collected at Arecibo Observatory. The Arecibo Observatory is part of the National Astronomy and Ionosphere Center, which is operated by Cornell University under a cooperative agreement with the National Science Foundation.\\ This research has made use of the NASA/IPAC Extragalactic Database (NED) which is operated by the Jet Propulsion Laboratory, California Institute of Technology, under contract with the National Aeronautics and Space Administration.\\
    Funding for the SDSS and SDSS-II has been provided by the Alfred P. Sloan Foundation, the Participating Institutions, the National Science Foundation, the U.S. Department of Energy, the National Aeronautics and Space Administration, the Japanese Monbukagakusho, the Max Planck Society, and the Higher Education Funding Council for England. The SDSS Web Site is http://www.sdss.org/.\\
    The SDSS is managed by the Astrophysical Research Consortium for the Participating Institutions. The Participating Institutions are the American Museum of Natural History, Astrophysical Institute Potsdam, University of Basel, University of Cambridge, Case Western Reserve University, University of Chicago, Drexel University, Fermilab, the Institute for Advanced Study, the Japan Participation Group, Johns Hopkins University, the Joint Institute for Nuclear Astrophysics, the Kavli Institute for Particle Astrophysics and Cosmology, the Korean Scientist Group, the Chinese Academy of Sciences (LAMOST), Los Alamos National Laboratory, the Max-Planck-Institute for Astronomy (MPIA), the Max-Planck-Institute for Astrophysics (MPA), New Mexico State University, Ohio State University, University of Pittsburgh, University of Portsmouth, Princeton University, the United States Naval Observatory, and the University of Washington.
\end{acknowledgements}

\bibliographystyle{aa} 
\bibliography{astroph} 

\end{document}